\documentclass[12pt]{article}
\usepackage{graphicx,psfrag,epsf}
\usepackage{enumerate}
\usepackage{url} % not crucial - just used below for the URL
\RequirePackage{amsthm,amsmath,amsfonts,amssymb}
\RequirePackage{natbib}

\usepackage{subfigure}
\usepackage{caption,booktabs,array}
\usepackage{tikz}
\usepackage{xcolor}
\usetikzlibrary{arrows.meta, positioning}
\usepackage{float}

\usepackage[colorlinks=true, citecolor=blue, urlcolor=blue, allcolors=blue]{hyperref}
\usepackage{listings}

\lstdefinelanguage{Stan}{
  keywords={data, parameters, model, transformed, real, int, vector, matrix},
  keywordstyle=\color{blue}\bfseries,
  comment=[l]{//},
  commentstyle=\color{gray},
  morecomment=[s]{/*}{*/},
  stringstyle=\color{red},
  sensitive=true
}

\lstdefinelanguage{R}{
  keywords={if, else, repeat, while, function, for, in, next, break},
  otherkeywords={TRUE, FALSE, NULL, NA, NaN, Inf},
  sensitive=true,
  comment=[l]{\#},
  string=[b]",
}

\newcommand{\blind}{1}

\usepackage[margin=1in]{geometry}

\begin{document}

\def\spacingset#1{\renewcommand{\baselinestretch}%
{#1}\small\normalsize} \spacingset{1}

%%%%%%%%%%%%%%%%%%%%%%%%%%%%%%%%%%%%%%%%%%%%%%%%%%%%%%%%%%%%%%%%%%%%%%%%%%%%%%

\if1\blind
{
  \title{\bf Modeling bounded well-being indices using Bayesian double generalized beta regression with spatial and temporal borrowing}
  \author{Abhi Jain, Michael LaValley, Kimberly A. Dukes, Shariq Mohammed\thanks{corresponding author}\hspace{.2cm}\\
    Department of Biostatistics, Boston University School of Public Health}
  \maketitle
} \fi

\if0\blind
{
  \bigskip
  \bigskip
  \bigskip
  \begin{center}
    {\LARGE\bf Modeling bounded well-being indices using Bayesian double generalized beta regression with spatial and temporal borrowing}
\end{center}
  \medskip
} \fi

\bigskip
\begin{abstract}
Health and well-being indices are widely used to assess population health outcomes and inform policy decisions. Individual-level assessment of well-being can be used to develop community-level indices that measure wellness for different geographical units. While many existing indices operate at coarse geographic levels such as counties or states, finer spatial resolution can offer more actionable insights. We present a novel Bayesian double generalized beta regression framework to model a bounded individual-level well-being index (WBI) using annual survey data collected from 2021 to 2023 in Massachusetts. Although survey respondents may differ across years, responses are geotagged to ZIP Code Tabulation Areas (ZCTAs), enabling the integration of both spatial and temporal information. Our framework incorporates spatial dependencies via a graph Laplacian matrix that encodes driving time-based ZCTA neighborhood structure, and leverages temporal borrowing by using posterior spatial effect estimates from one year to inform priors in the next. This dual-borrowing strategy within a Bayesian double generalized beta regression framework enhances estimation precision, particularly in areas with sparse data, and improves inference for smaller geographic units. We demonstrate the utility of our method through a realistic simulation study that highlights improved estimation when borrowing spatial and temporal information. In the real data analysis, we model individual-level well-being for residents in Massachusetts and find that income, education status, and marital status are most associated with WBI. Additionally, we observe that ZCTAs in Western Massachusetts, Cape Cod, and those near Boston perform best with the highest spatial effects. 
\end{abstract}

\noindent%

\vfill

\newpage
\spacingset{1.45} % DON'T change the spacing!
\section{Introduction}

Quantifying health and well-being at fine geographic scales is essential for identifying community-specific challenges and informing targeted policy interventions. Several publicly available indices have been developed to measure health and well-being outcomes across different geographic levels. For example, the Robert Wood Johnson Foundation's County Health Rankings produce a composite score based on over 30 health outcomes and health factor measures to rank counties nationwide \citep{remington2015county1}. The Gallup-Healthways Well-Being Index surveys approximately 500 adults daily and aggregates responses across five main dimensions--physical, community, financial, social, and purpose--at the state or metropolitan statistical area (MSA) level \citep{skopec2014potential, wood2015financial}. Other indices, such as the Whole Health Index \citep{chi2023whole} and the Environmental Protection Agency's Human Well-Being Index \cite{summers2017development} also operate primarily at the county level. However, the AARP's Livability Index provides scores at more granular spatial units, including Census block groups \citep{guzman2015livability}.

Many of these indices, including the well-being index (WBI) analyzed in this study, are defined on a bounded scale. However, few modeling frameworks explicitly account for this constraint. Beta regression models are well-suited for such data. While commonly applied to model rates and proportions \citep{ferrari2004beta, douma2019analysing}, beta regression has also been used to analyze bounded scores and indices. For example, \citet{hunger2012longitudinal} employed a longitudinal beta regression model to analyze health-related quality of life scores, and \citet{rogers2012combining} developed a beta regression framework to model cognitive scores in Alzheimer's disease research. A key advantage of the beta distribution is its two-parameter structure, which allows for flexible modeling of both the mean and precision (dispersion). Double generalized beta regression--also referred to as varying dispersion beta regression--models both the mean and variance jointly as a function of external covariates \citep{simas2010improved,cepeda2015beta,zhao2014variable} and was first proposed in a Bayesian context by \citet{cepeda2001variability}. These models have been applied in educational research to model student performance in mathematics, language and reading ability \citep{cepeda2013spatial,bayer2017model}. 

In this paper, we propose a novel Bayesian double generalized beta regression framework to model individual-level WBI using annual survey data collected from 2021 to 2023 in Massachusetts. Although survey respondents vary across years, responses are geotagged to ZIP Code Tabulation Areas (ZCTAs), enabling the integration of spatial and temporal information.\footnote{ZCTAs are the approximations of ZIP Codes developed by the United States Census Bureau and used for mapping and statistics analysis \cite{bureau_2022}. ZCTAs offer finer spatial resolution and tend to be more homogeneous in demographic and socioeconomic characteristics \citep{grubesic2008zip}.} Existing applications of beta regression seldom incorporate spatial or temporal dependencies, which can be reasonable when working with large and heterogeneous spatial units such as states or counties. However, leveraging spatial structure becomes critical for improving estimation accuracy in settings involving smaller geographical units (e.g., ZCTAs) with limited data within them. In our framework, we incorporate spatial structure using a graph Laplacian matrix constructed from driving-time-based adjacency between ZCTA population centroids. \citet{cepeda2013spatial} incorporated spatial lags in a Bayesian double generalized beta regression model, and \citet{martinez2019hierarchical} used spatially structured random effects in point-referenced data. Our approach adapts these ideas to areal data at the ZCTA level. Additionally, within our double generalized modeling setup, we model the mean with individual-level covariates and the dispersion with ZCTA-level covariates, allowing us to capture both within- and between-ZCTA variability in well-being. Our goal here is to capture characteristics of ZCTAs that could inform levels of inequality or disparities of well-being within a ZCTA. Additionally, we have have developed software to implement our model, which can be found at \url{https://github.com/abhijainstats/BayesBadger}.

% As far as we can tell based on available documentation, the aforementioned health and well-being indices  do not borrow information spatially through spatial smoothing techniques. This makes sense in large part because states and even counties are large spatial units with high degrees of heterogeneity within the unit. However, in our paper, we use ZIP Codes, or more specifically ZIP Code Tabulation Areas (ZCTAs), which are the approximations of ZIP Codes developed by the United States Census Bureau and used for mapping and statistics analysis \cite{bureau_2022}. ZCTAs which are much smaller spatial units and thought to be more homogeneous in terms of its demographic and socioeconomic characteristics \citep{grubesic2008zip}. Thus, within our double generalized Beta regression framework, we also employ a graph Laplacian matrix to inform spatial dependencies between ZCTAs. \citet{cepeda2013spatial} also incorporated spatial dependencies in their Bayesian double-generalized Beta regression through spatial lags, while \citet{martinez2019hierarchical} used spatially structured random effects, but in the context of point-referenced spatial data.

We fit separate Bayesian beta regression models to annual well-being survey data for three consecutive years. Borrowing temporal information across years is particularly valuable for small geographical units, where data can be sparse or highly heterogeneous. This approach stabilizes estimates by leveraging persistent spatial patterns and reduces overfitting by introducing regularization through prior-year information. Within a Bayesian framework, it also offers a principled way to incorporate prior knowledge across time. Specifically, we use posterior estimates of ZCTA-level spatial effects from one year as informative priors for the subsequent year--a strategy conceptually similar to Bayesian borrowing, which has been applied in clinical trial settings where data are often collected in batches \citep{muehlemann2023tutorial}. Our proposed framework incorporates (i) spatial information through a geographically informed neighborhood structure and (ii) temporal information by leveraging prior-year posterior estimates, which we collectively refer to as borrowing in the remainder of this paper.

To the best of our knowledge, this is the first application of a double generalized beta regression framework that jointly integrates both spatial and temporal dependencies, and has not been previously explored. Spatial dependencies are informed through a graph Laplacian, in which neighbors are defined based on driving times between ZCTA population centroids, which is distinct from more standard spatial smoothing techniques like conditional autoregressive (CAR) models \citep{besag1974spatial}. Using driving times between ZCTA population centroids to inform neighbors gives a more accurate reflection of individual's utilization of resources and amenities outside their own ZCTA. Specifically, we use a 30-minute threshold to define neighbors as individuals on average commute 28 minutes one-way for work \citep{burd2021travel} and 94\% of Americans live with a 30-minute drive of HRSA-supported health center \citep{rankin2024quantifying}. Another advantage of using a graph Laplacian to inform the spatial dependency is that our model can estimate ZCTA-level effects for ZCTAs with no observations by borrowing information from neighboring ZCTAs. Furthermore, by using a Bayesian framework, the model is also able to estimate ZCTA-level effects for ZCTAs with no observations and no neighbors by sampling from the prior, whose covariance is already spatially informed. In addition to borrowing information spatially, we also borrow temporally from data from previous years. By using posterior estimates for ZCTA-level effects from the previous year's model as the prior for the subsequent year, we are implementing a form of temporal regularization, which is realistic as we would not expect ZCTA-level effects to differ significantly year-to-year. This temporal borrowing approach also differs from standard time-series models, such as an AR(1) model, which requires a sufficient amount of time points to reliably estimate the autocorrelation. In contrast, our approach encodes temporal dependence directly through the prior, enabling information to be shared over time, without requiring many time points, which is particularly advantageous in our case with only three years of survey data.

To evaluate our method, we conduct a simulation study demonstrating the benefits of borrowing spatial and temporal information. We then apply our model to real-world well-being data from Massachusetts, identifying ZCTAs in Western Massachusetts, Cape Cod and Boston suburbs as the ones with the highest ZCTA spatial effects. Additionally, from our model, we find that income, marital status, and higher education have the largest association with well-being. The rest of the paper is organized as follows: Section \ref{sec:bbregression} outlines the modeling framework and estimation approach;
Section \ref{sec:simulation} presents the simulation study; Section \ref{sec:casestudy} introduces the data and details the results from real data analysis; and Section \ref{sec:discussion} concludes with a discussion on findings, limitations, and future directions.

\section{Bayesian beta regression with borrowing}\label{sec:bbregression}

The outcome of interest, WBI, is bounded on the interval $[0,100]$. Since the support for the beta distribution is $(0,1)$, we transform WBI scores accordingly. A simple rescaling by dividing by 100 would suffice if the data excluded boundary values. However, since WBI scores can take values exactly equal to 0 or 100, we apply the transformation proposed by \cite{smithson2006better} to ensure all transformed values lie strictly within $(0,1)$:
\[Y = \frac{Y'(N-1)+0.5}{N},\] where $Y'=WBI/100$ and $N$ is the sample size. This transformation preserves the relative ordering of the data while ensuring compatibility with the beta distribution.

\subsection{Beta distribution.} We assume a beta distribution for the transformed WBI score $Y$. Density function for the beta distribution under standard parameterization is given by
\begin{equation}
    f(y; a, b) = \frac{\Gamma(a + b)}{\Gamma(a) \Gamma(b)} y^{a-1} (1-y)^{b-1},~~~~~ y \in (0,1),
    \label{eqn:beta1}
\end{equation}
where $a > 0$ and $b > 0$ are shape parameters. Following the reparameterization proposed by \citet{ferrari2004beta}, we express the density in Equation \ref{eqn:beta1} in terms of mean $\mu = \frac{a}{a+b}$ and precision parameter $\phi = a + b$, yielding:
\begin{equation}
    f(y;\mu, \phi) = \frac{\Gamma(\phi)}{\Gamma(\mu \phi) \Gamma((1-\mu)\phi)} y^{\mu \phi -1} (1-y)^{(1-\mu)\phi-1}, ~~~~~0 < \mu < 1, \phi > 0.
    \label{eqn:beta2}
\end{equation}
The parameterization is particularly useful in regression settings as $E[Y] = \mu$ and $Var(Y) = \frac{\mu(1-\mu)}{1+\phi}$. While $\phi$ is not exactly a precision parameter since $\phi \neq Var(Y)^{-1}$, $\phi$ can be interpreted as precision parameter since $Var(Y)$ increases as $\phi$ decreases, assuming that $\mu$ is held constant \citep{ferrari2004beta}. In what follows, we adopt this parameterization to flexibly model both the mean and precision of the WBI scores as functions of covariates.

\subsection{Model Specification}\label{subsec:model}
Let $t=1, \ldots, T$ index survey years, and let $s = 1, \ldots, S$ denote ZCTAs. For each year $t$, we observe $n_{st}$ respondents in ZCTA $s$, with a total of $n_t=\sum_{s=1}^S n_{st}$ respondents. Let $Y_{ist} \in (0,1)$ represent the transformed WBI score for individual $i$ from ZCTA $s$ during year $t$, where $Y_{ist}$'s are assumed independent across $i$, $s$, and $t$. This assumption is reasonable for our data since each year's survey data is a repeated cross-section, with minimal number of participants responding in multiple years. We assume that \( Y_{ist} \) follows a beta distribution with mean \( \mu_{ist} \) and precision \( \phi_{st} \), parameterized as:
%Since we employ a Bayesian modeling framework with priors dependent on previous year's estimates, we will first examine the model setup for the initial year $t=1$ and then subsequent years $t=2, \ldots, T$.
%For year $t$, we model $y_{ist}$ (WBI) using a Beta distribution as shown below:
\begin{equation}
    f(y_{ist};\mu_{ist}, \phi_{st}) = \frac{\Gamma(\phi_{st})}{\Gamma(\mu_{ist} \phi_{st}) \Gamma((1-\mu_{ist})\phi_{st})} y^{\mu_{ist} \phi_{st} -1} (1-y)^{(1-\mu_{ist})\phi_{st}-1},
    \label{eqn:betaModel}
\end{equation}
where $\mu_{ist}$ and $\phi_{ist}$ are modeled using external covariates through link functions as described next. We model the mean response \( \mu_{ist} \) as:

\begin{equation}
    g(\mu_{ist}) = \eta_{ist} = \mathbf{x}_{ist}^\top \boldsymbol{\beta}_t + \alpha_{st} = \alpha_0 + \mathbf{x}_{ist}^\top \boldsymbol{\beta}_t + \tilde{\alpha}_{st},
    \label{eqn:linkFunMu}
\end{equation}
where $g$ is a link function that maps the real line to $(0,1)$, $\mathbf{x}_{ist}$ is a $p \times 1$ vector of individual-level covariates for participant $i$ in ZCTA $s$ at year $t$, $\boldsymbol{\beta}_t$ is $p \times 1$ vector of corresponding regression coefficients for year $t$, and $\alpha_{st}$ is the ZCTA-level spatial effect for ZCTA $s$ at year $t$. Several different link functions can be considered, such as probit or log-log, but we select the logit link function, which is the most commonly used link function for beta regression. The spatial effect $\alpha_{st}$ can be decomposed into a global average ZCTA-level effect $\alpha_0$ and a centered ZCTA-level spatial deviation $\tilde{\alpha}_{st}$ such that $\alpha_{st} = \alpha_0 + \tilde{\alpha}_{st}$ and $\sum_{s=1}^S \tilde{\alpha}_{st} = 0$. The precision parameter $\phi_{st}$ is modeled as:

\begin{equation}
    h(\phi_{st}) = \zeta_{st} = \omega_0 + \mathbf{c}_{st}^\top \boldsymbol{\omega}_t,
    \label{eqn:linkFunPhi}    
\end{equation}
where $h$ is a logarithmic link function, $\omega_0$ is the intercept, $\mathbf{c}_{st}$ is a $q \times 1$ vector of ZCTA-level covariates for ZCTA $s$ at year $t$, and $\boldsymbol{\omega}_t$ is $q \times 1$ vector of corresponding regression coefficients for year $t$.

Thus, for year $t=1$, the hierarchical spatial Bayesian double generalized beta regression model that combines the mean and precision models can be written as:
\begin{align*}
Y_{ist} \mid \boldsymbol{\beta}_t, \alpha_{st}, \mathbf{x}_{ist}, \boldsymbol{\omega}_t, \mathbf{c}_{st} & \sim \text{Beta}(\mu_{ist}, \phi_{st}) \equiv f(y|\boldsymbol{\beta}, \boldsymbol{\alpha}, \mathbf{x}, \boldsymbol{\omega}, \mathbf{c}),
\end{align*}
with the following priors on the model parameters:
\begin{align*}
\boldsymbol{\beta}_t & \sim N_p(\textbf{0}, \tau_{\mu} I_p) \equiv \pi(\boldsymbol{\beta}) \\
\boldsymbol{\alpha}_t & \sim N_S(\textbf{0}, M^{-1}) \equiv \pi(\boldsymbol{\alpha}|\lambda, \gamma) \\
\boldsymbol{\omega}_t & \sim N_q(\textbf{0}, \tau_{\phi} I_q) \equiv \pi(\boldsymbol{\omega}) \\
\lambda & \sim Gamma(\theta_1, \theta_2) \equiv \pi(\lambda) \\
\gamma & \sim Gamma(\iota_1, \iota_2) \equiv \pi(\gamma),
\end{align*}
where $\tau_{\mu}$ and $\tau_{\phi}$ are hyperparameters informing the variance of the regression coefficients for the mean and precision models, respectively. A table detailing all the model parameters and their definitions is provided in the Supplementary Materials.

% Additionally, $M= \lambda L + \lambda\gamma I_S$, where $L = D-A \in \mathbb{R}^{S \times S}$ is the graph Laplacian matrix constructed from the ZCTA adjacency structure, with $D$ and $A$ denoting the degree and adjacency matrices, respectively. The two tuning parameters $\lambda >0$ and $\gamma >0$ control the spatial smoothness and numerical stability of the ZCTA-level effects. 

\subsection{Spatial and Temporal Borrowing}\label{subsec:spatial}

\subsubsection{Spatial smoothing via the graph Laplacian.} Spatial information is incorporated through the prior covariance of the ZCTA-level effects $\boldsymbol{\alpha_t}$, specified as $M^{-1}$, where \( M = \lambda L + \lambda \gamma I_S \) is a function of the graph Laplacian matrix $L = D-A \in \mathbb{R}^{S \times S}$, with $D$ and $A$ denoting the degree and adjacency matrices, respectively. Additionally, the two tuning parameters $\lambda >0$ and $\gamma >0$ control the spatial smoothness and numerical stability of the ZCTA-level effects. The weights of the adjacency matrix are defined as $A_{ss'} = \min\left\{\exp\left(-\frac{\delta_{ss'}^2}{2\rho^2}\right), \exp\left(-\frac{\delta_{s's}^2}{2\rho^2}\right)\right\}$, where $\delta_{ss'}$ and \( \delta_{s's} \) denote the driving times from \( s \) to \( s' \) and from \( s' \) to \( s \), respectively, and $\rho$ is the driving time threshold (set to 30 minutes in our application). Since driving times are determined by road networks, traffic conditions, and other factors, the driving time from ZCTA $s$ to $s'$ is not necessarily equivalent to the driving time from ZCTA $s'$ to $s$. However, our definition ensures symmetry in \( A \), which is required for the multivariate normal prior. For example, if the driving time from $s$ to $s'$ is 31 minutes but only 29 minutes from $s'$ to $s$, the two ZCTAs are still considered neighbors. ZCTAs whose population centroids are more than 30 minutes away from one another are not considered neighbors and assigned a weight of zero. The degree matrix \( D \) is diagonal, with entries $D_{ss} = \sum_{s'=1}^S A_{ss'}$ and $D_{ss'} = 0$ for \( s \neq s' \). 

\subsubsection{Temporally informed spatial effects.} For years $t \geq 2$, we model $y_{ist}$ using the same hierarchical Bayesian double generalized beta regression framework described above. The key difference lies in the prior specification for the spatial effects $\boldsymbol{\alpha}_t$. For year $t \geq 2$, instead of using a multivariate normal prior with mean zero, we incorporate information from previous year(s) using posterior estimates of spatial effects $\{\widehat{\boldsymbol{\alpha}}_t \mid t \in \{1,\ldots,t-1\}\}$. That is, the prior distribution for year $t$ can be expressed as:
$$\boldsymbol{\alpha}_t \sim N_s(p_1\widehat{\boldsymbol{\alpha}}_{t-1}+p_2\widehat{\boldsymbol{\alpha}}_{t-2} + \ldots + p_{t-1}\widehat{\boldsymbol{\alpha}}_{1} + q \mathbf{0}, M^{-1}),$$
where $\widehat{\boldsymbol{\alpha}}_{t-j}$ denotes the posterior estimate of $\boldsymbol{\alpha}$ from year $t-j$ and 
$p_1, \ldots,p_k$ are weights. This structure enables the model to adaptively borrow strength across time while maintaining flexibility in prior informativeness. Some special cases of this prior include: (i) a mean zero prior when when $p_1, \ldots, p_k = 0$ and $q=1$; (ii) an informative prior based solely on the previous year when $p_1=1$ and all other waits are zero; and (iii) a hybrid prior that combines multiple years of information with decay and shrinkage over time, that is, $p_1>p_2> \ldots >p_{t-1}$ and/or $p_1, \ldots, p_l \in (0,1)$ for $l < t-1$ and $p_{l+1}, \ldots, p_{t-1}, \ldots = 0$.  %This would state that last year's posteriors estimates are more influential that the posterior estimates from two year ago and so on. Lastly, by setting $0<q<1$, we can shrink posterior estimates from previous years' towards zero so that the prior specification is a linear combination of an informative component and non-informative component.
%However, sensitivity analysis is performed using other choices for the prior of $\boldsymbol{\alpha}_t$ and can be found in the Appendix.
In this study, we only borrow information from the previous year for simplicity and due to the limited number of years available. Specifically, we set $p_1=1$ and $p_2=\ldots=p_{t-1}=q=0$ resulting in the prior distribution for $\boldsymbol{\alpha}_t$ as $N_S(\widehat{\boldsymbol{\alpha}}_{t-1}, M^{-1})$. The posterior distribution and parameter estimation remain the same as described below. %and the posterior distribution of the model for year $t \ge 2$ can be expressed as:
% \begin{equation}
% \pi(\boldsymbol{\beta}, \boldsymbol{\alpha}, \boldsymbol{\omega}, \lambda, \gamma | y) \propto f(y|\boldsymbol{\beta}, \alpha, \boldsymbol{\omega}, \lambda, \gamma) \pi(\boldsymbol{\beta}) \pi_2(\boldsymbol{\alpha} | \lambda, \gamma) \pi(\boldsymbol{\omega}) \pi(\lambda) \pi(\gamma)
% \label{eqn:posterior2}
% \end{equation}

\subsection{Posterior Inference and Marginal Effects}\label{subsec:posterior}

\subsubsection{Parameter estimation.} The joint posterior distribution for the parameters in year \( t \) is given as:
\[
\pi(\boldsymbol{\beta}, \boldsymbol{\alpha}, \boldsymbol{\omega}, \lambda, \gamma \mid \mathbf{y}) \propto 
\prod_{s=1}^{S} \prod_{i=1}^{n_{st}} f(y_{ist} \mid \mu_{ist}, \phi_{st}) \cdot 
\pi(\boldsymbol{\beta}) \pi(\boldsymbol{\alpha} \mid \lambda, \gamma) \pi(\boldsymbol{\omega}) \pi(\lambda) \pi(\gamma).
\]
The full joint posterior distribution and the full conditional distribution for the spatial effects $\boldsymbol{\alpha}$ are provided in the Supplementary Materials, along with derivations of all other full conditionals. Note that, the full conditionals--particularly for the spatial effects $\boldsymbol{\alpha}$--are analytically intractable and do not follow known family of distributions. Hence, to perform posterior inference, we employ the Stan probabilistic programming language \citep{carpenter2017stan}. Stan implements Hamiltonian Monte Carlo and is well-suited for high-dimensional models with complex posterior geometries, offering efficient exploration of the parameter space and faster convergence compared to traditional Markov chain Monte Carlo (MCMC) sampling methods.

\subsubsection{Marginal effects}\label{subsubsec:marginal}
Because overall WBI is modeled on the logit scale, the interpretation of regression coefficients is not straightforward. Although the outcome is a continuous score bounded between 0 and 1, it is not a proportion in the traditional sense. To facilitate interpretation, we compute marginal effects by transforming the linear predictor using the inverse logit (expit) function. For a given year $t$, the expected value of \( Y_{is} \) is:%$logit(\mu_{is}) = \mathbf{x}_{is} \boldsymbol{\beta} + \alpha_{s}$ and that $E[y_{is}] = \mu_{is}$. Thus, by applying the expit function, we can express $E[y_{is}]$ as:
\begin{equation}
    E[Y_{is}] = \frac{\text{exp}(\mathbf{x}_{is} \widehat{\boldsymbol{\beta}} + \widehat{\alpha}_{s})}{1+\text{exp}(\mathbf{x}_{is} \widehat{\boldsymbol{\beta}} + \widehat{\alpha}_{s})},
    \label{eqn:expY}
\end{equation}
where $\mathbf{x}_{is}$ is a vector of covariates for individual $i$ in ZCTA $s$, $\widehat{\boldsymbol{\beta}}$ are the estimated regression coefficients, and $\widehat{\alpha}_{s}$ is the estimated ZCTA-level effect. Using Equation \ref{eqn:expY}, we compute average marginal effects  for continuous, binary, and categorical covariates. %to more clearly interpret our regression coefficients and ZCTA-level effects. Below we present derivation of the marginal effects for continuous, binary, and categorical variables.
This formulation of the marginal effects allows for interpretation on the original scale of the WBI scores, ranging from 0 to 100, by multiplying the expected values by 100, while also noting that the outcome $Y_{is}$ has been transformed slightly to ensure that all values lie between 0 and 1 exclusive.

\paragraph{Continuous variables} For year $t$, consider the covariates $X_1, \ldots, X_p$ for the mean model. Let $X_1$ be the continuous covariate of interest. The average marginal effect \( \delta_c \) of a one-unit increase in \( X_1 \) is computed as:
\begin{align*}
    \delta_c &= E[Y | X_1 = x_1+1, X_2 = x_2, \ldots, X_p = x_p, \alpha] - E[Y | X_1 = x_1, X_2 = x_2, \ldots, X_p = x_p, \alpha] \\
    &= \sum_{s=1}^S \sum_{s:i=1}^{n_s} \frac{e^{(x_{1is}+1) \hat{\beta}_1 + x_{2is} \hat{\beta}_2 + \ldots x_{pis} \hat{\beta}_p + \hat{\alpha}_s}}{1+e^{(x_{1is}+1) \hat{\beta}_1 + x_{2is} \hat{\beta}_2 + \ldots x_{pis} \hat{\beta}_p + \hat{\alpha}_s}} - \frac{e^{x_{1is} \hat{\beta}_1 + x_{2is} \hat{\beta}_2 + \ldots x_{pis} \hat{\beta}_p + \hat{\alpha}_s}}{1+e^{x_{1is} \hat{\beta}_1 + x_{2is} \hat{\beta}_2 + \ldots x_{pis} \hat{\beta}_p + \hat{\alpha}_s}}.
\end{align*}
Thus, $\delta_c$ represents the average change in the expected WBI score associated with a one unit increase in $X_1$, holding all other covariates constant. To quantify uncertainty, we compute the marginal effect \( \delta_c \) at each iteration of the MCMC sampling procedure. This yields a posterior distribution for the average marginal effect, from which we can derive point estimates and corresponding credible intervals.

\paragraph{Binary variables} For a binary covariate \( X_1 \in \{0, 1\} \), the average marginal effect \( \delta_b \) is defined as:
%Let's assume that for year $t \in \mathcal{T}$, there are $p$ variables $X_1, \ldots, X_p$ that are used to model the mean $\mu$. We are interested in the effect of a one unit change in $X_1$ (a binary variable) on the outcome. We can then compute the average marginal effect $\delta_b$ of $X_1$ on $Y$ with the following calculation:
\begin{align*}
    \delta_b &= E[Y | X_1 = 1, X_2 = x_2, \ldots, X_p = x_p, \alpha] - E[Y | X_1 = 0, X_2 = x_2, \ldots, X_p = x_p, \alpha] \\
    &= \sum_{s=1}^S \sum_{s:i=1}^{n_s} \frac{e^{\hat{\beta}_1 + x_{2is} \hat{\beta}_2 + \ldots x_{pis} \hat{\beta}_p + \hat{\alpha}_s}}{1+e^{\hat{\beta}_1 + x_{2is} \hat{\beta}_2 + \ldots x_{pis} \hat{\beta}_p + \hat{\alpha}_s}} - \frac{e^{x_{2is} \hat{\beta}_2 + \ldots x_{pis} \hat{\beta}_p + \hat{\alpha}_s}}{1+e^{x_{2is} \hat{\beta}_2 + \ldots x_{pis} \hat{\beta}_p + \hat{\alpha}_s}}.
\end{align*}
Thus, $\delta_b$ represents the average difference in expected WBI scores between individuals with \( X_1 = 1 \) and those with \( X_1 = 0 \), conditional on other covariates.

\paragraph{Categorical variables} A similar calculation can be used to find the average marginal effect for categorical variables. Consider an example where 
%there are $p$ variables $X_1, \ldots, X_p$ that are used to model the mean $\mu$ and 
$X_1$ is a categorical variable with $K$ levels and suppose $X_1 \in \{1,2,\ldots,K\}$ with $X_1=1$ as the reference level. Then the average marginal effect for level $k ~(\neq 1)$ relative to level 1 is:
\begin{align*}
    \delta_f &= E[Y | I(X_1 = k), X_2 = x_2, \ldots, X_p = x_p, \alpha] - E[Y | I(X_1=1), X_2 = x_2, \ldots, X_p = x_p, \alpha] \\
    &= \sum_{s=1}^S \sum_{s:i=1}^{n_s} \frac{e^{\hat{\beta}_1^{(k)} + x_{2is} \hat{\beta}_2 + \ldots x_{pis} \hat{\beta}_p + \hat{\alpha}_s}}{1+e^{\hat{\beta}_1^{(k)} + x_{2is} \hat{\beta}_2 + \ldots x_{pis} \hat{\beta}_p + \hat{\alpha}_s}} - \frac{e^{x_{2is} \hat{\beta}_2 + \ldots x_{pis} \hat{\beta}_p + \hat{\alpha}_s}}{1+e^{x_{2is} \hat{\beta}_2 + \ldots x_{pis} \hat{\beta}_p + \hat{\alpha}_s}},
\end{align*}
where $\hat{\beta}_1^{(k)}$ is the regression coefficient for the indicator variable $I(X_1=k)$ and since $X_1=1$ is the reference level, the coefficient for $I(X_1=1)$ is zero. Similar expressions can be derived for other levels. These marginal effects provide interpretable summaries of the impact of categorical predictors on the expected WBI score.

\section{Simulation study}\label{sec:simulation}

The simulation study is designed to evaluate three key aspects of out modeling framework. First, we assess the model's ability to accurately capture the effects of covariates in both the mean and precision models. Second, we examine the accuracy with which the model estimates ZCTA-level spatial effects. Third, we investigate whether borrowing information from year to year--implemented by using posterior estimates of ZCTA effects from one year as priors for the next--improves estimation accuracy. Simulations are conducted under four distinct spatial patterns and three precision settings.

\subsection{Simulation Setup}\label{subsec:sim_setup}

\begin{figure}[!t]
\centering
\begin{tabular}{|c|c|}
\hline
\subfigure[Block]{\label{fig:block_true}\includegraphics[trim=0cm 2cm 0cm 2cm, clip, scale=0.3, page=1]{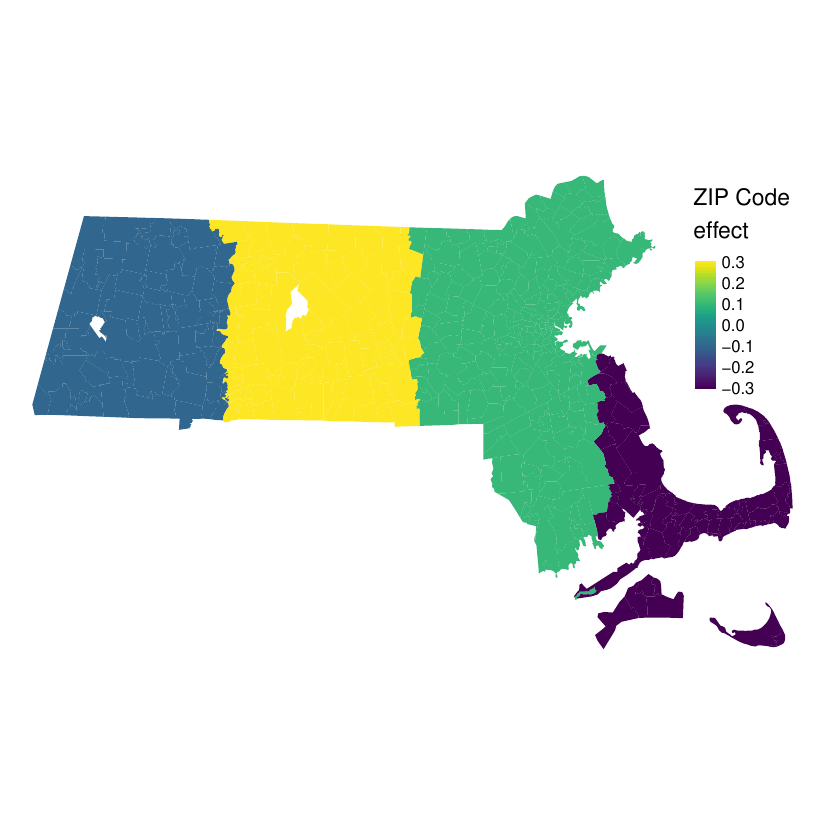}} & 
\subfigure[Smooth]{\label{fig:smooth_true}\includegraphics[trim=0cm 2cm 0cm 2cm, clip, scale=0.3, page=2]{plots_sim_trueAlphaChoropleths.pdf}} \\ \hline
\subfigure[Hotspots]{\label{fig:hotspots_true}\includegraphics[trim=0cm 2cm 0cm 2cm, clip, scale=0.3, page=3]{plots_sim_trueAlphaChoropleths.pdf}} &
\subfigure[Random]{\label{fig:random_true}\includegraphics[trim=0cm 2cm 0cm 2cm, clip, scale=0.3, page=4]{plots_sim_trueAlphaChoropleths.pdf}} \\ \hline
\end{tabular}
% }
\caption{True ZCTA-level spatial effects under four simulation scenarios: (a) Block, (b) Smooth, (c) Hotspots, and (d) Random. }
\end{figure}

We conduct a realistic simulation study by using actual individual-level survey data and ZCTA-level demographic covariates, while simulating only the outcome variable. The number of observations matches that of the Massachusetts Sharecare data for each of the three years 2021-2023. We consider four spatial patterns \citep{halder2021spatial}--\textit{block}, \textit{smooth}, \textit{hotspots}, and \textit{random}--illustrated in Figures \ref{fig:block_true}-\ref{fig:random_true}. 

In the block pattern, ZCTAs are grouped into one four equal-sized vertical blocks based on latitudes of the ZCTA population centroids. However, due to the shape of Massachusetts, we assign ZCTAs on the upper right of the state to Block 3 instead of Block 4 to preserve spatial continuity. %The true spatial effects is determined by their block assignment and range from $-$0.3 to 0.3, though blocks are non-consecutive and 
The true spatial effects for Blocks 1 through 4 are set to $-0.1$, $0.3$, $0.1$, and $-0.3$, respectively. In the smooth pattern, the true spatial effects are set to increase continuously from west to east, ranging from $-0.3$ to $0.3$. The \textit{hotspots} pattern assigns positive and negative spatial effects centered around Boston and Springfield, respectively, with true effects decaying as a function of distance from these cities. Lastly, in the random pattern, the true spatial effects are randomly sampled from a uniform distribution $\mathcal{U}{\{-3,3}\}$. 

%In our double generalized Beta regression framework, we use the individual-level covariates for the mean model and MA ZIP Code level demographic variables to model the precision. 
For the mean model, we use three individual-level covariates: a binary indicator variable for female sex, a standardized continuous age variable (\( X_2 \)), and a categorical income variable (\( X_3 \)) with four levels $(1=\text{under }25\text{K}, 2=25\text{-}50\text{K}, 3=50\text{-}100\text{K}, 4=\text{over }100\text{K})$, with under $25$K as the reference category. The design matrix includes: $\{X_1, X_2, I(X_3=2), I(X_3=3), I(X_3=4)\}$. The linear predictor for the mean is specified as:
$$\text{logit}(\mu_{ist}) = \alpha_0 + \beta_1 X_1 + \beta_2 X_2 + \beta_3 (X_3=2) + \beta_4 (X_3=3) + \beta_5 (X_3=4) + \tilde{\alpha}_{st},$$
where $\alpha_0 = 0.1$, $\beta_1 = -0.1$, $\beta_2 = 0.4$, $\beta_3 = 0.2$, $\beta_4 = 0.4$, $\beta_5 = 0.6$. The ZCTA-level effect $\tilde{\alpha}_{st}$ is defined by the spatial pattern and the overall spatial effect is given by $\alpha_{st} = (\alpha_0 +\tilde\alpha_{st})$.
%\frac{\mu_{ist}}{1-\mu_{ist}} = \eta_{ist} 

For the initial year, $\tilde{\alpha}_{st}$ is prespecified and defined by the spatial pattern under consideration as described above. To simulate temporal variation in spatial effects, we perturb the previous-year values using a multivariate normal distribution: $
\tilde{\alpha}_{st} = \tilde{\alpha}_{s,t-1} + \epsilon_{st}, ~\epsilon_{st} \sim \mathcal{N}(\mathbf{0}, \mathbf{\Sigma}) \text{ for } t \geq 2,
$
%However, in order to emulate a more realistic scenario, we add a small perturbation $\epsilon_{st}$ to $\tilde{\alpha}_{st}$ in subsequent years. We simulate this perturbation from a $\mathcal{N}(\mathbf{0}, \mathbf{\Sigma})$ distribution, 
where $\mathbf{\Sigma} = 0.005(\lambda \mathbf{L} + \lambda \gamma \mathbf{I})^{-1}$, with $\lambda=10$ and $\gamma=0.1$ representing the expected values of the spatial smoothing parameters. %To test the expected standard deviation and range of perturbations from this distribution, we sampled 1000 times from $\mathcal{N}(\mathbf{0}, \mathbf{\Sigma})$ and found that 
Based on 1,000 draws from this distribution, the average standard deviation of the perturbations is approximately $0.009$, with a typical range of $(-0.06, 0.06)$. %Thus, in subsequent years the true ZIP Code effect is $\tilde{\alpha}_{st} = \tilde{\alpha}_{s,t-1} +\epsilon_{st}$.

For the precision model, we use three ZCTA-level covariates: $C_1$ (proportion of residents in the ZCTA with a college degree), $C_2$ (proportion of White residents), and $C_3$ (proportion of residents aged 65 and above). Although the covariates used to model precision in this simulation are not direct measures of inequality or disparity, their range $[0,1]$ matches that of the diversity/segregation measures used in the case study using real data from Massachusetts.  %Each of these variables are are continuous and bounded from 0 to 1. 
Thus, we simulate the precision $\phi$, as:
$$log(\phi_{st}) = \zeta_{st} = \omega_0 + \omega_1 C_1 + \omega_2 C_2 + \omega_3 C_3,$$ where $\omega_0$ represents the intercept and can take on three values $(\omega_0 \in \{1, 3, 5\})$ corresponds to \textit{low}, \textit{medium}, and \textit{high} precision settings, and $\omega_1=1.5, \omega_2=0.5,$ and $\omega_3=-0.8$. The outcome is then simulated as: $Y_{ist} \sim Beta(\mu_{ist}, \phi_{st})$. 

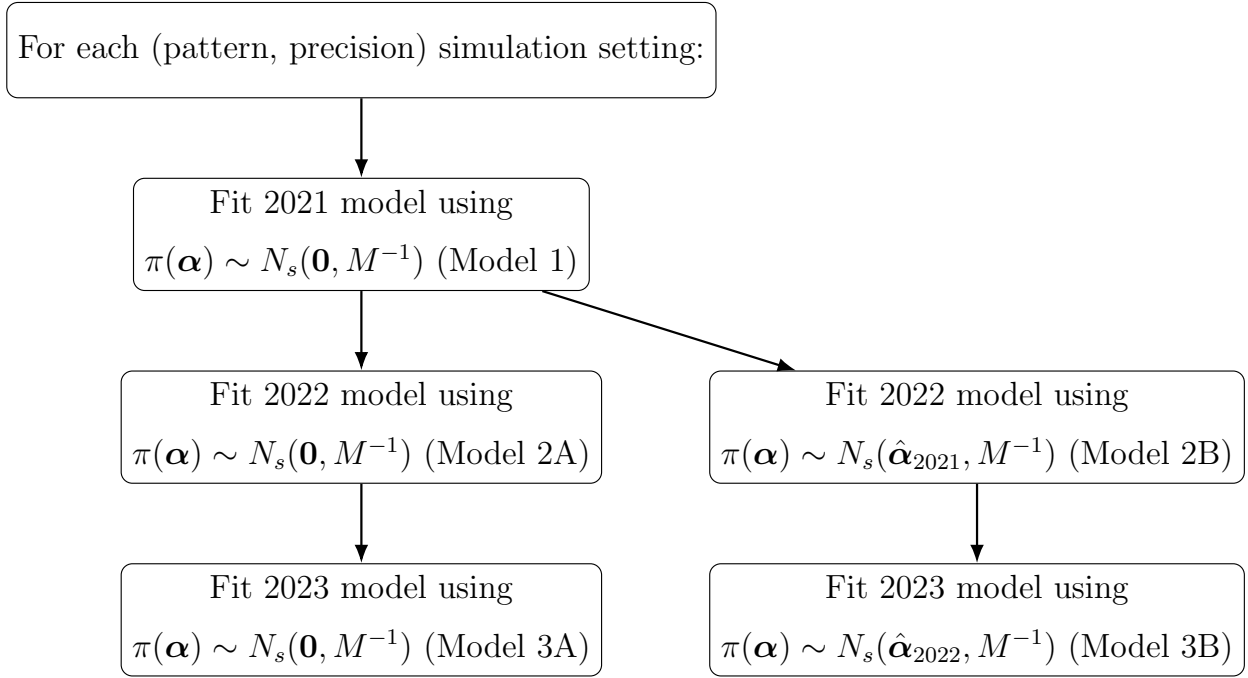
\begin{figure}
\resizebox{\textwidth}{!}{%
\begin{tikzpicture}[
  node distance = 1cm and 1.5cm,
  every node/.style = {draw, rectangle, rounded corners,
                       align=center, minimum width=3.5cm,
                       minimum height=1.2cm},
  arrow/.style  = {-{Latex}, thick}
]
% --- nodes -------------------------------------------------
\node (start)      {For each (pattern, precision) simulation setting:};
\node (fit2021)  [below=of start]
                   {Fit 2021 model using \\
                   $\pi(\boldsymbol{\alpha})\sim N_s(\mathbf{0}, M^{-1})$ (Model 1)};
\node (fitNI2022) [below=of fit2021]
                   {Fit 2022 model using \\
                   $\pi(\boldsymbol{\alpha})\sim N_s(\mathbf{0}, M^{-1})$ (Model 2A)};
\node (fitI2022)  [below right=of fit2021]
                   {Fit 2022 model using \\
                   $\pi(\boldsymbol{\alpha})\sim N_s(\hat{\boldsymbol{\alpha}}_{2021}, M^{-1})$ (Model 2B)};
\node (fitNI2023) [below=of fitNI2022]
                   {Fit 2023 model using \\$\pi(\boldsymbol{\alpha})\sim N_s(\mathbf{0}, M^{-1})$ (Model 3A)};
\node (fitI2023)  [below=of fitI2022]
                   {Fit 2023 model using \\$\pi(\boldsymbol{\alpha})\sim N_s(\hat{\boldsymbol{\alpha}}_{2022}, M^{-1})$ (Model 3B)};
% --- arrows -----------------------------------------------
\draw[arrow] (start)        -- (fit2021);
\draw[arrow] (fit2021)      -- (fitNI2022);
\draw[arrow] (fit2021)    -- (fitI2022);
\draw[arrow] (fitNI2022)     -- (fitNI2023);
\draw[arrow] (fitI2022) -- (fitI2023);
\end{tikzpicture}
}
\caption{Flow chart of the models that are run for the simulation study. For each (pattern, precision) simulation setting, we perform 30 replications where the above models are run for each replication. To evaluate whether temporal borrowing aids estimation of ZCTA-level effects, we compare Model 2A with Model 2B and Model 3A with Model 3B.\label{fig:flow}}
\end{figure}

Overall, there are 12 total simulation settings (four spatial pattern settings and three precision settings). We perform 30 replications for each simulation setting, where the models are fit for each year 2021-2023 for each replication, as illustrated in Figure \ref{fig:flow}.

\subsection{Simulation results}
As mentioned above, we evaluate model performance across three key aspects: (1) recovery of regression coefficients in the mean and precision models, (2) accuracy of ZIP Code-level spatial effect estimation, and (3) the impact of temporal borrowing on spatial estimation.

\subsubsection{Regression coefficients} 
%We first examine whether true regression coefficients in both the mean and precision models are accurately captured. 
Table \ref{tab:sim_coef} presents the 95\% credible interval coverage rates for regression coefficients under the block spatial pattern across three precision settings, three years, and two prior specifications. Our model consistently recovers the true coefficients in the mean model, with coverage exceeding 90\% in most scenarios. %we can see that for the block spatial pattern, the true beta coefficients in the mean model are being captured in over 90\% of the 95\% credible intervals for almost all precision settings and across all years and prior specifications. Slightly lower coverage is observed in the 2022 model with an informative prior for $\beta_3 - \beta_5$, which are the three levels for the Income categorical variable, 
Notably, coverage for the income-related coefficients (\( \beta_3, \beta_4, \beta_5 \)) dips slightly in 2022 under the informative prior but returns to above 90\% in 2023. In contrast, coverage in the precision model is more variable. Under low and medium precision, coverage in 2021 ranges from 70\% to 90\%, improving in subsequent years. However, under high precision, coverage is poor in 2021, especially for the precision model intercept $(\omega_0)$ and the coefficient for proportion of residents with a college degree $(\omega_1)$. Coverage does improve to over 90\% in later years. Additionally, while coverage is poor in 2021, upon further examination, the 95\% credible intervals for $\omega_0$ and $\omega_1$ are not substantially different from the true values, with the largest deviation for $\omega_0$ across the 30 replicates being $(5.18, 5.59)$ and the largest deviation for $\omega_1$ being $(-0.04, 1.05)$. These results suggest that the model either requires more information to reliably estimate precision parameters or might not require modeling of precision separately, particularly in high-precision settings due to minimal variability in within-ZCTA overall WBI. The results for the smooth and hotspots patterns are similar to those observed under the block pattern. Notably, the extremely low coverage rates seen in the block pattern under the high precision setting do not occur in the smooth or hotspots patterns. The lowest coverage observed in these settings was 70\% and 73\%, respectively, both corresponding to precision model parameters in the initial year under high precision. For the random spatial pattern, the model performed well in capturing the mean model coefficients. However, as with the other spatial patterns, it struggled to accurately estimate precision model parameters under high precision setting in all years. Full results for the smooth, hotspots, and random patterns are provided in the Supplementary Materials. 

\begin{table}[!t]
\caption{\label{tab:sim_coef} 95\% credible interval coverage (in percent) of mean and precision model parameters for the Block pattern across precision settings, years, and prior specifications. ``NI'' = non-informative prior; ``I'' = informative prior using previous year's posterior estimates.}
\resizebox{\textwidth}{!}{%
\small
\begin{tabular}{c|ccccc|ccccc|ccccc|}
\cline{2-16}
& \multicolumn{5}{c|}{Low Precision} & \multicolumn{5}{c|}{Medium Precision} & \multicolumn{5}{c|}{High Precision} \\
\cline{2-16}
& \multicolumn{1}{c|}{2021} & \multicolumn{2}{c|}{2022} & \multicolumn{2}{c|}{2023}
& \multicolumn{1}{c|}{2021} & \multicolumn{2}{c|}{2022} & \multicolumn{2}{c|}{2023}
& \multicolumn{1}{c|}{2021} & \multicolumn{2}{c|}{2022} & \multicolumn{2}{c|}{2023} \\
\hline
\multicolumn{1}{|c|}{Parameter} & \multicolumn{1}{c|}{NI} & NI & \multicolumn{1}{c|}{I} & NI & I
& \multicolumn{1}{c|}{NI} & NI & \multicolumn{1}{c|}{I} & NI & I
& \multicolumn{1}{c|}{NI} & NI & \multicolumn{1}{c|}{I} & NI & I \\ \hline
\multicolumn{1}{|c|}{$\beta_1$} & \multicolumn{1}{c|}{90} & 90 & \multicolumn{1}{c|}{90} & 100 & 100 & \multicolumn{1}{c|}{97} & 100 & \multicolumn{1}{c|}{97} & 93 & 93 & \multicolumn{1}{c|}{97} & 90 & \multicolumn{1}{c|}{93} & 100 & 97 \\
\multicolumn{1}{|c|}{$\beta_2$} & \multicolumn{1}{c|}{100} & 97 & \multicolumn{1}{c|}{97} & 90 & 90 & \multicolumn{1}{c|}{93} & 93 & \multicolumn{1}{c|}{93} & 93 & 97 & \multicolumn{1}{c|}{90} & 97 & \multicolumn{1}{c|}{97} & 90 & 90 \\
\multicolumn{1}{|c|}{$\beta_3$} & \multicolumn{1}{c|}{100} & 83 & \multicolumn{1}{c|}{80} & 97 & 90 & \multicolumn{1}{c|}{90} & 90 & \multicolumn{1}{c|}{90} & 97 & 97 & \multicolumn{1}{c|}{97} & 90 & \multicolumn{1}{c|}{90} & 93 & 93 \\
\multicolumn{1}{|c|}{$\beta_4$} & \multicolumn{1}{c|}{100} & 90 & \multicolumn{1}{c|}{77} & 87 & 90 & \multicolumn{1}{c|}{87} & 93 & \multicolumn{1}{c|}{90} & 100 & 93 & \multicolumn{1}{c|}{90} & 93 & \multicolumn{1}{c|}{97} & 97 & 97 \\
\multicolumn{1}{|c|}{$\beta_5$} & \multicolumn{1}{c|}{93} & 90 & \multicolumn{1}{c|}{87} & 93 & 93 & \multicolumn{1}{c|}{93} & 100 & \multicolumn{1}{c|}{93} & 97 & 97 & \multicolumn{1}{c|}{97} & 93 & \multicolumn{1}{c|}{100} & 93 & 97 \\ \hline
\multicolumn{1}{|c|}{$\omega_0$} & \multicolumn{1}{c|}{80} & 100 & \multicolumn{1}{c|}{100} & 97 & 97 & \multicolumn{1}{c|}{83} & 93 & \multicolumn{1}{c|}{93} & 100 & 100 & \multicolumn{1}{c|}{40} & 93 & \multicolumn{1}{c|}{93} & 93 & 97 \\
\multicolumn{1}{|c|}{$\omega_1$} & \multicolumn{1}{c|}{90} & 97 & \multicolumn{1}{c|}{97} & 97 & 93 & \multicolumn{1}{c|}{77} & 100 & \multicolumn{1}{c|}{100} & 93 & 93 & \multicolumn{1}{c|}{30} & 90 & \multicolumn{1}{c|}{83} & 90 & 90 \\
\multicolumn{1}{|c|}{$\omega_2$} & \multicolumn{1}{c|}{70} & 93 & \multicolumn{1}{c|}{93} & 90 & 90 & \multicolumn{1}{c|}{83} & 100 & \multicolumn{1}{c|}{100} & 90 & 93 & \multicolumn{1}{c|}{73} & 83 & \multicolumn{1}{c|}{83} & 90 & 93 \\
\multicolumn{1}{|c|}{$\omega_3$} & \multicolumn{1}{c|}{80} & 97 & \multicolumn{1}{c|}{97} & 93 & 93 & \multicolumn{1}{c|}{90} & 83 & \multicolumn{1}{c|}{90} & 93 & 93 & \multicolumn{1}{c|}{93} & 80 & \multicolumn{1}{c|}{87} & 90 & 93 \\
\hline
\end{tabular}
}
\end{table}

%Table \ref{tab:sim_rmse} reports the root mean squared error (RMSE) of ZCTA effect estimates across all spatial patterns. The model performs best under the smooth spatial pattern, with RMSE values as low as $0.041$ in the high precision setting and around $0.075$ in the low precision setting. Since the true spatial effects range from $-0.3$ to $0.3$, these estimates are reasonably accurate as they may be off by less than $0.075$ and $0.04$ units in the low and high precision settings, respectively. For the block and hotspots patterns, RMSEs range from about 0.135 in the low precision setting to 0.08 and 0.11 in the high precision setting. Unsurprisingly, the higher RMSE values were observed for the random pattern, with values ranging from 0.175 in the low precision setting to 0.11 in the high precision setting, reflecting the lack of spatial structure. %Even though there is some room for improvement in terms of the ZIP Code effect estimation. Notably, the overall spatial pattern was preserved for each setting.

\subsubsection{Spatial effects} 
\begin{figure}[!t]
\centering
\begin{tabular}{|c|c|}
\hline
\subfigure[Block]{\label{fig:block_est}\includegraphics[trim=2cm 0cm 2cm 0cm, clip, scale=0.24, page=1]{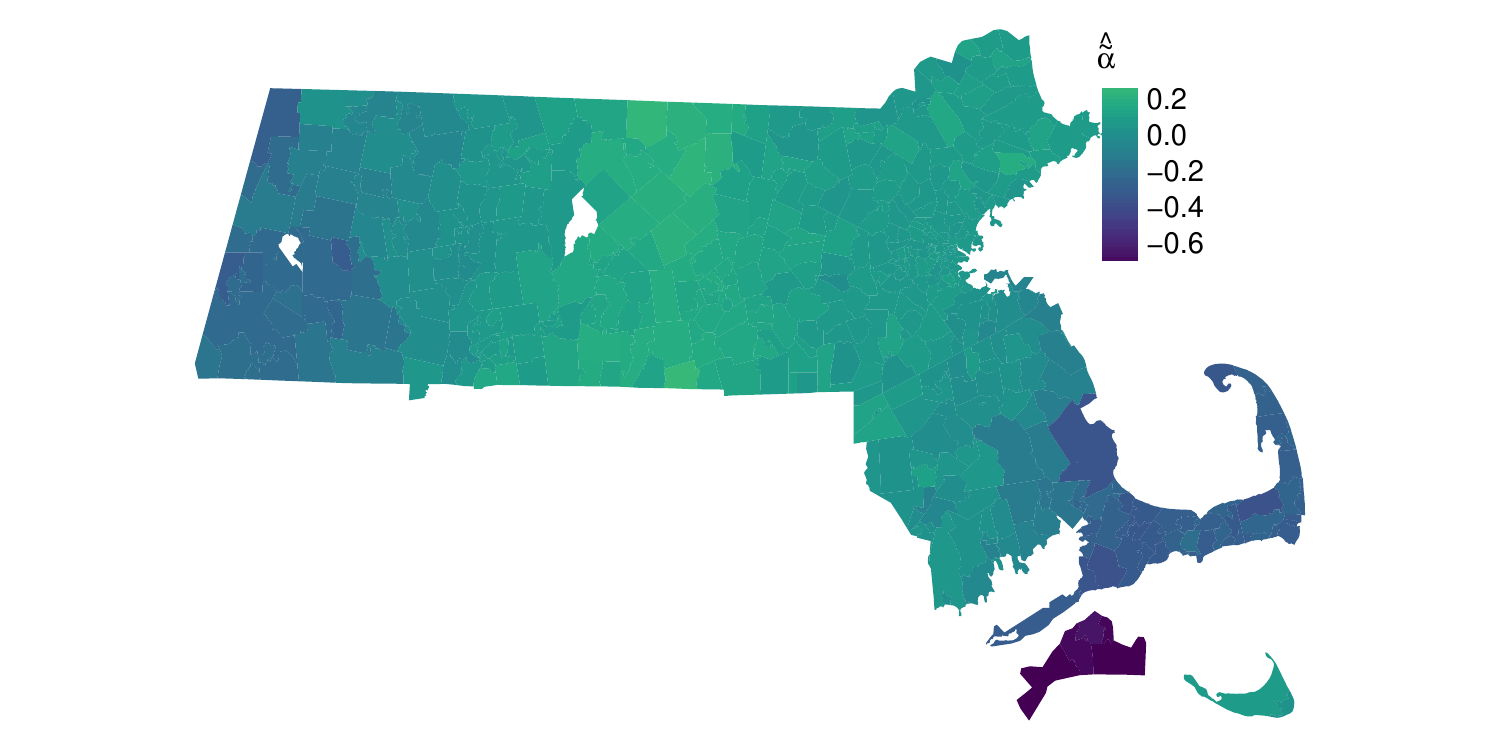}} & 
\subfigure[Smooth]{\label{fig:smooth_est}\includegraphics[trim=2cm 0cm 2cm 0cm, clip, scale=0.24, page=2]{plots_sim_zcta-choropleths-sample-2.pdf}} \\ \hline
\subfigure[Hotspots]{\label{fig:hotspots_est}\includegraphics[trim=2cm 0cm 2cm 0cm, clip, scale=0.24, page=3]{plots_sim_zcta-choropleths-sample-2.pdf}} &
\subfigure[Random]{\label{fig:random_est}\includegraphics[trim=2cm 0cm 2cm 0cm, clip, scale=0.24, page=4]{plots_sim_zcta-choropleths-sample-2.pdf}} \\ \hline
\end{tabular}
\caption{Choropleth maps of estimated ZIP Code effects for a single replicate under the low precision setting across the four spatial patterns for the year 2021.}
\end{figure}

Figure~\ref{fig:block_est}–\ref{fig:random_est} show estimated ZCTA-level effects for a single replicate under the low precision setting. The choropleths for other replications illustrate similar spatial distributions of ZCTA-level effects. The estimated spatial patterns closely resemble the true configurations. For example, in the block pattern, the ZIP Codes in central Massachusetts have the highest $\tilde{\hat{\alpha}}$ estimates and Cape Code, Martha's Vineyard and Nantucket have the lowest $\tilde{\hat{\alpha}}$ estimates (Figure \ref{fig:block_est}). In the smooth pattern, the true pattern of ZCTA spatial effects increasing as we move eastwards is preserved. In the hotspots pattern, we observe the lower $\tilde{\hat{\alpha}}$ estimates near Springfield with higher estimates near the Boston area, consistent with the true pattern (Figure \ref{fig:hotspots_est}). Additionally, Table 4 in the Supplementary Materials reports the average root mean squared error (RMSE) of ZCTA effect estimates across all replications for all spatial patterns and precision settings. We find that the model performs best at capturing the true spatial effects under the smooth spatial pattern and worst under the random pattern, with the block and hotspots patterns' performance in between.

\subsubsection{Impact of temporal borrowing} We assess the benefit of using posterior estimates from the previous year as priors for ZIP Code effects. Figure~\ref{fig:spearman} displays the average Spearman correlation between true and estimated ZIP Code effects across 30 replicates. Overall, temporal borrowing improves estimation in low precision settings and does not degrade performance in medium and high precision settings. In low precision settings, an informative prior leads to higher Spearman correlation between $\tilde{\alpha}$ and $\tilde{\hat{\alpha}}$ across the replications in 2022 and 2023 for the block, smooth, and hotspots patterns, with the difference being statistically significant for the block and hotspots patterns. In medium and high precision settings, differences between prior specifications are minimal. For the random pattern, the Spearman correlation remains low ($0.2$ to $0.25$), with only modest improvement in 2023 under the informative prior. Thus, there is evidence to suggest that while borrowing information temporally may not always yield better estimation, it is helpful in settings with low precision and not harmful in other settings. We also compare our proposed model with a non-spatial frequentist beta regression model with random effects fit using the \texttt{glmmTMB} R package \citep{brooks2017glmmtmb} and more standard Bayesian beta regression with spatial random effects using an ICAR prior fit using the \texttt{brms} R package \citep{burkner2017brms}. As shown in Figure \ref{fig:spearman}, the ICAR model underperforms both graph Laplacian models for the block spatial pattern, and it also yields poorer results for the smooth and hotspot patterns under the low-precision setting. Moreover, its performance further degrades in subsequent years, as the ICAR model does not employ any temporal borrowing mechanism. The non-spatial model performs very poorly, as expected, in the block, smooth, and hotspots settings, but does outperform the other three methods in the random pattern setting. Furthermore, both the non-spatial and the ICAR models are unable to capture the ZCTA-level effects of ZCTAs with no observations from standard model implementations, whereas the proposed model with graph Laplacian spatial smoothing allows for estimation of ZCTA-level effects for ZCTAs with no observations, so long as there is a neighboring ZCTA with at least one observation. We omitted results of the non-spatial model from Figure \ref{fig:spearman} for visual clarity, though we have included the same figure with all four methods in the Supplementary materials.

\begin{figure}[!t]
\centering
\includegraphics[scale=.57]{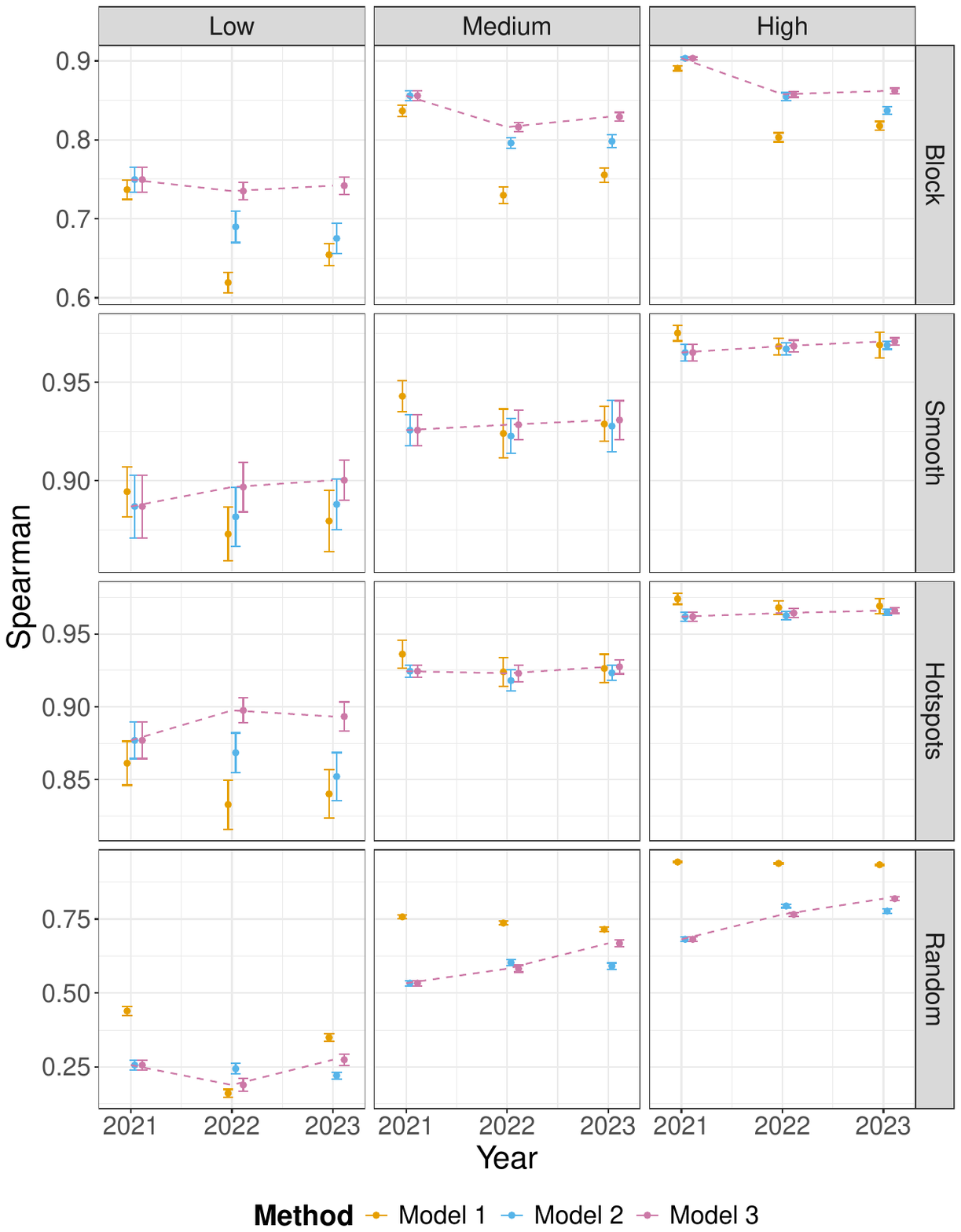}
\caption{\label{fig:spearman}Mean spearman correlation between true and estimated ZIP Code effects across spatial patterns, precision settings, and years. Error bars denote 95\% confidence intervals across the 30 replicates. Model 1 represents the Bayesian beta regression with spatial random effects using an ICAR prior. Model 2 represents the proposed model without temporal borrowing and Model 3 represents the proposed model with temporal borrowing.}
\end{figure}

\section{Case study: Massachusetts}\label{sec:casestudy}

\subsection{Data description}\label{subsec:data}

We use individual-level survey data collected by Sharecare, a digital health company that administers health and well-being assessments across the United States. The surveys evaluate  five key dimensions of well-being: physical, financial, social, community, and purpose. We will use the \textit{overall WBI}--a score between 0 to 100--as the primary outcome variable in our analysis. Sharecare collects approximately 500,000 responses annually; for this study, we focus on annual survey data conducted in Massachusetts from 2021 to 2023. 

\subsubsection{Coverage and WBI summary statistics} 
\begin{figure}[!t]
\centering
\begin{tabular}{|c|c|c|}
\hline
\subfigure[2021]{\label{fig:coverage21}\includegraphics[trim=0cm 2cm 0cm 2cm, clip, scale=0.23, page=1]{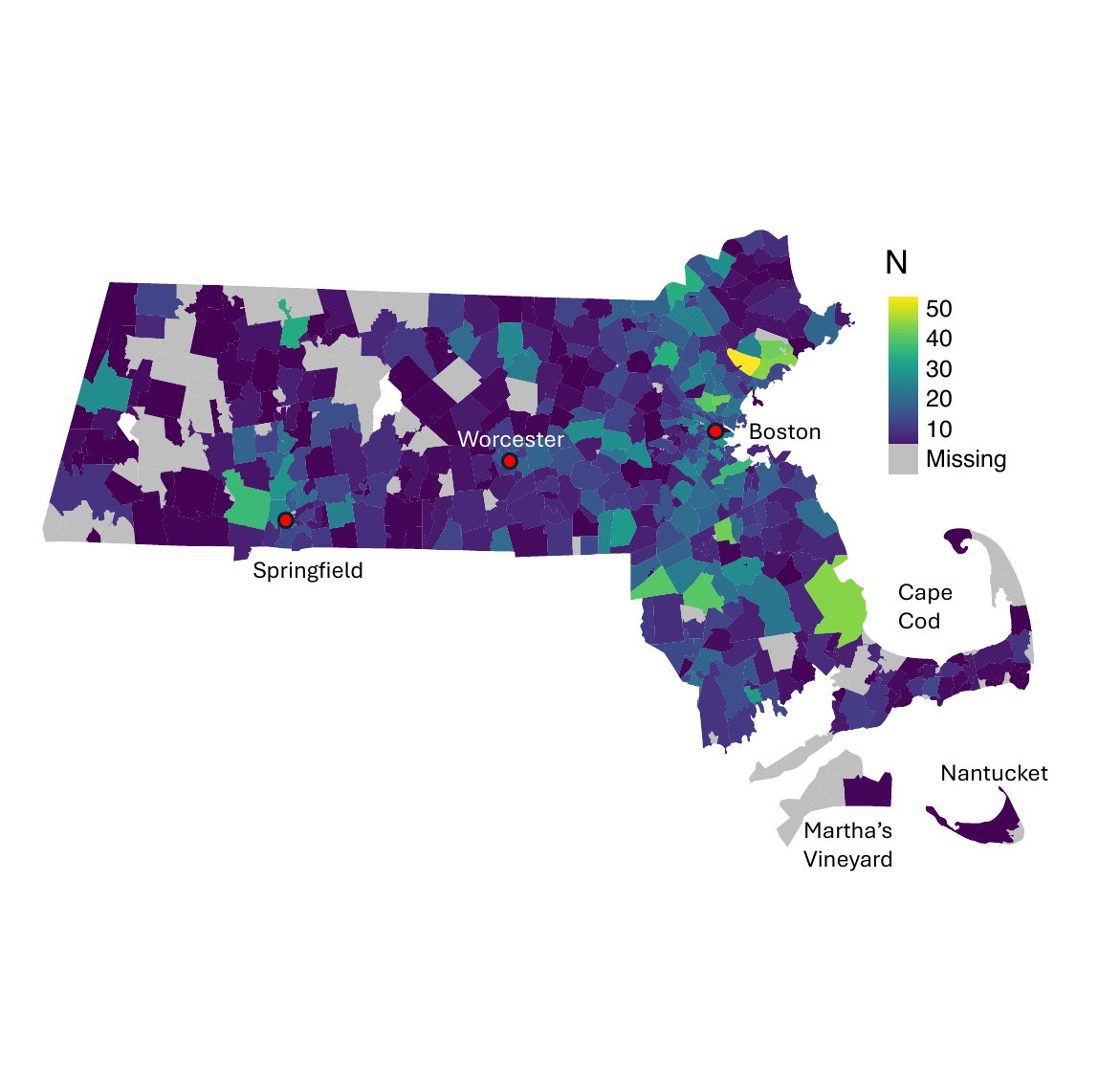}} & 
\subfigure[2022]{\label{fig:coverage22}\includegraphics[trim=0cm 2cm 0cm 2cm, clip, scale=0.36, page=2]{plots_data_zctaCoverageMaps.pdf}} &
\subfigure[2023]{\label{fig:coverage23}\includegraphics[trim=0cm 2cm 0cm 2cm, clip, scale=0.36, page=3]{plots_data_zctaCoverageMaps.pdf}} \\ \hline
\end{tabular}
\caption{Geographic coverage of survey respondents across ZCTAs in Massachusetts for the years 2021–2023. Lighter shades indicate higher response counts, while darker shades represent lower response count. ZCTAs shaded gray are ones with zero responses.}
\end{figure}

\begin{figure}[!h]
\centering
\begin{tabular}{|c|c|c|}
\hline
\subfigure[2021]{\label{fig:avgWbi2021}\includegraphics[trim=0cm 2cm 0cm 2cm, clip, scale=0.3, page=1]{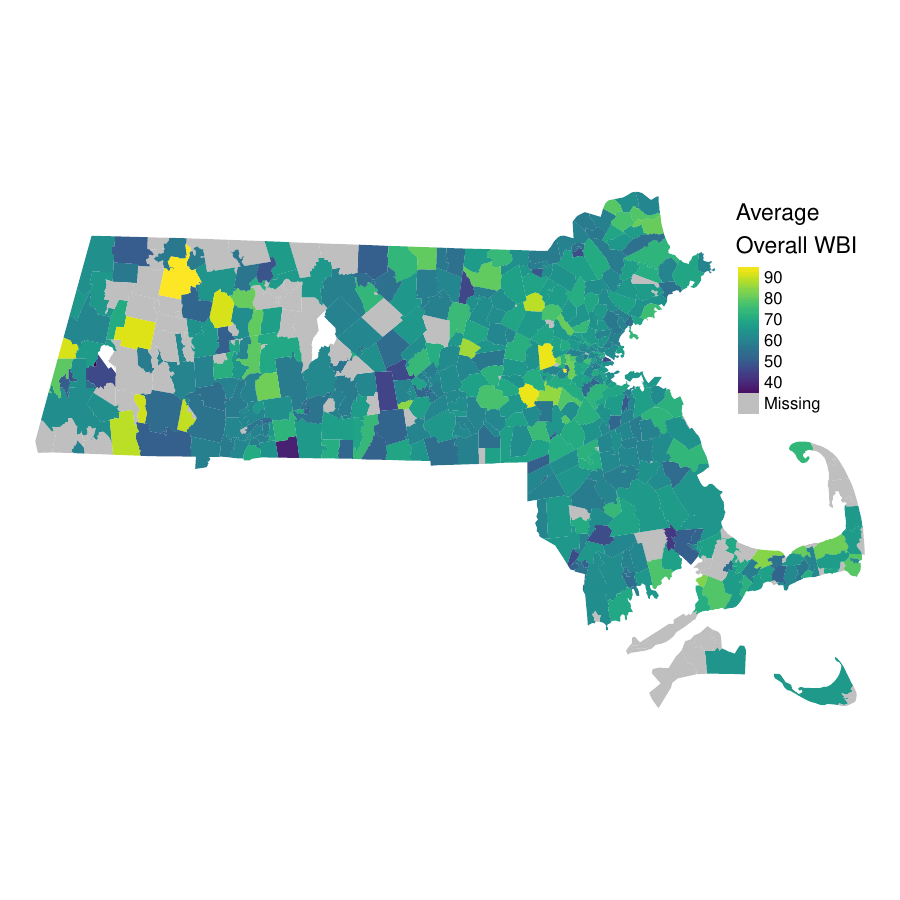}} & 
\subfigure[2022]{\label{fig:avgWbi2022}\includegraphics[trim=0cm 2cm 0cm 2cm, clip, scale=0.3, page=2]{plots_data_avgWbiByYear.pdf}} &
\subfigure[2023]{\label{fig:avgWbi2023}\includegraphics[trim=0cm 2cm 0cm 2cm, clip, scale=0.3, page=3]{plots_data_avgWbiByYear.pdf}} \\ \hline
\end{tabular}
\caption{Choropleth maps illustrating the average overall WBI by ZCTA for the years 2021, 2022, and 2023. Darker shades indicate lower average WBI scores, while lighter shades represent higher scores. ZCTAs shaded gray are ones with zero responses.}
\end{figure}

The Sharecare surveys are administered continuously throughout the year and compiled into annual datasets. In Massachusetts, the sample includes 4,443 respondents in 2021, 3,602 in 2022, and 3,428 in 2023. These are repeated cross-sectional surveys with minimal overlap in respondents across years. Figures \ref{fig:coverage21} - \ref{fig:coverage23} display the geographic coverage by ZCTA for each year. Massachusetts is comprised of 537 ZCTAs. In 2021, survey responses were recorded in 448 ZCTAs, with slightly less coverage in subsequent years---439 ZCTAs in 2022 and 435 in 2023. Across all three years, coverage is highest in Eastern Massachusetts, especially in and around Greater Boston area. In contrast, Western Massachusetts and parts of Cape Cod were consistently underrepresented in the sample. The average WBI increased slightly over time, from 63.2 in 2021 to 64.2 in 2023, with a standard deviation around 16 in each year. Figure~\ref{fig:avgWbi2021}--\ref{fig:avgWbi2023} show the spatial distribution of average overall WBI by ZCTA for each year.

\begin{table}[!t]
\centering
\caption{\label{tab:demo_by_year} Demographic characteristics among Massachusetts survey respondents, by year (2021–2023). Values represent the percentage of respondents within each demographic category for each survey year.}

% Adjust width to fit your document margins
\begin{minipage}{7.5cm}
\begin{tabular}{@{}lrrr@{}}
\toprule
\textbf{Variable} & \textbf{2021} & \textbf{2022} & \textbf{2023} \\
\midrule
\textbf{Gender}\\
\hspace{4mm}Male & 38.7 & 45.1 & 56.0 \\
\hspace{4mm}Female & 61.3 & 54.9 & 44.0 \\
\textbf{Age}\\
\hspace{4mm}18--29 & 24.4 & 23.8 & 25.2 \\ 
\hspace{4mm}30--44 & 29.8 & 29.3 & 30.3 \\ 
\hspace{4mm}45--64 & 33.6 & 31.5 & 29.3 \\ 
\hspace{4mm}65+ & 12.2 & 15.3 & 15.1 \\ 
\textbf{Race}\\
\hspace{4mm}White & 82.2 & 73.9 & 66.2 \\ 
\hspace{4mm}Black & 5.5 & 7.3 & 13.0 \\ 
\hspace{4mm}Hispanic/Latino & 5.6 & 10.6 & 12.2 \\ 
\hspace{4mm}Asian & 5.0 & 6.5 & 6.9 \\ 
\hspace{4mm}Other & 1.8 & 1.7 & 1.7 \\
\bottomrule
\end{tabular}
\end{minipage}
\hspace{0.5cm} % spacing between the tables
\begin{minipage}{5cm}
\begin{tabular}{@{}lrrr@{}}
\toprule
\textbf{Variable} & \textbf{2021} & \textbf{2022} & \textbf{2023} \\
\midrule
\textbf{Marital Status}\\
\hspace{4mm}Never Married & 37.5 & 33.7 & 37.3 \\  
\hspace{4mm}Married & 48.5 & 52.8 & 50.4 \\ 
\hspace{4mm}Divorced & 14.0 & 13.5 & 12.3 \\
\textbf{Education}\\
\hspace{4mm}$<$High School & 2.9 & 1.7 & 3.6 \\  
\hspace{4mm}High School & 24.7 & 22.8 & 22.8 \\ 
\hspace{4mm}College & 52.6 & 52.6 & 50.8 \\
\hspace{4mm}Post-Graduate & 19.8 & 22.9 & 22.8 \\ 
\textbf{Income}\\
\hspace{4mm}$<$25K & 14.6 & 12.0 & 13.7 \\ 
\hspace{4mm}25K--50K & 18.6 & 16.1 & 14.7 \\ 
\hspace{4mm}50K--100K & 30.4 & 31.0 & 30.1 \\ 
\hspace{4mm}100K+ & 36.4 & 40.9 & 41.5\\ 
\bottomrule
\end{tabular}
\end{minipage}
\end{table}

\subsubsection{Sample characteristics} Respondents are recruited through four channels: employer-based outreach (the majority), Sharecare’s mobile app, community-level surveys, and targeted surveys to reach underrepresented populations. Table \ref{tab:demo_by_year} shows the demographic breakdown of Massachusetts sample by year. The sample is largely comprised of highly educated and high-income individuals, with over 73\% of respondents holding at least a Bachelor's degree and more than 36\% reporting annual incomes above \$100,000. The racial composition of the sample became more diverse over time, with the proportion of White respondents changing from over 82\% in 2021 to 66\% in 2023. The proportion of male respondents also increased, from about 39\% to 56\%.

\begin{table}[!h]
\centering
\caption{\label{tab:zcta-vars} Definitions and summary statistics for ZCTA-level indices, including gender composition, absolute measures of age diversity, economic and educational segregation, and racial diversity.}
\begin{tabular}{lccc}
\toprule
\textbf{Index} & \textbf{Definition} & \textbf{Mean (SD)} & \textbf{Median (IQR)} \\
\midrule
\% Female & $\frac{\text{\# Female}}{\text{Population}}$ & 0.51 (0.07) & 0.51 (0.49, 0.53) \\
$\left|\text{Age Diversity}\right|$ & $\left|\frac{(\%\,{>65}) - (\%\,{<18})}{(\%\,{>65}) + (\%\,{<18})}\right|$ & 0.24 (0.23) & 0.18 (0.08, 0.31) \\
$\left|\text{Econ. Segregation}\right|$ & $\left|\frac{(\%\,{>100K}) - (\%\,{<25K})}{(\%\,{>100K}) + (\%\,{<25K})}\right|$ & 0.58 (0.25) & 0.62 (0.41, 0.77) \\
$\left|\text{Educ. Segregation}\right|$ & $\left|\frac{(\%\,{>BA}) - (\%\,{<HS})}{(\%\,{>BA}) + (\%\,{<HS})}\right|$ & 0.72 (0.25) & 0.80 (0.59, 0.91) \\
Racial Diversity & $1 - \sum_{g=1}^{8} p_g^2$ & 0.29 (0.19) & 0.25 (0.14, 0.43) \\
\bottomrule
\end{tabular}
\end{table}

\subsubsection{ZCTA-level covariates} To account for variation in well-being within ZCTAs, we incorporate ZCTA-level covariates derived from American Community Survey (ACS) measurements to model the precision parameter through the double generalized beta regression framework. Specifically, we consider variables that measure diversity and socioeconomic inequality that could potentially be associated with variation in well-being within ZCTAs. Definitions and summary statistics for these variables are presented in Table \ref{tab:zcta-vars}. The absolute age diversity index is a function of the difference in proportions of seniors and minors and thus a greater value indicates a larger imbalance between the groups. Similarly, absolute economic and educational segregation indices measure disparities between high- and low-income groups and college-educated versus high-school dropout populations, respectively. These variables are also known as index of concentration at the extremes \citep{tabb2024spatial}. The racial diversity index, obtained from the United States Census Bureau, is a function of the proportion of eight different non-overlapping race categories observed within a ZCTA \citep{census2021measuring}, where $p_g$ in Table \ref{tab:zcta-vars} represents the proportion belonging to each group. A larger racial diversity index corresponds to a more racially heterogeneous population.

\subsection{Modeling WBI in Massachusetts (2021--23)}\label{subsec:modeling}

\subsubsection{Model specification} We applied a Bayesian beta regression framework to model the overall WBI across Massachusetts for the three years 2021 through 2023. The outcome variable was individual-level overall WBI transformed within $(0,1)$ (Section \ref{sec:bbregression}). Individual-level covariates for the mean model include: sex, age, race, marital status, education, and income. The precision (dispersion) component was modeled using ZCTA-level covariates reflecting variability in different aspects within a ZCTA (Section \ref{subsec:data}). These included percentage female, age diversity, economic residential segregation, educational residential segregation, and racial diversity index. For the initial year (2021), we specified a mean-zero multivariate normal prior for the ZCTA-level spatial effect. In subsequent years (2022 and 2023), we used an informative prior structure for the spatial effects by using the posterior modes from the previous year as the prior means.

\subsubsection{Estimation strategy} Model estimation was conducted using Stan, with 10,000 posterior samples, a warm-up of 4,000 iterations, and thinning by a factor of 5. Convergence diagnostics, including trace plots and Gelman-Rubin statistics (all $<$1.012), indicated strong evidence of mixing and convergence for all model parameters in each year \citep{gelman1992inference}. Further details on model diagnostics can be found in the Supplementary materials.

\subsection{Parameter estimates and inference}\label{subsec:inference}

\begin{table}[!t]
\centering
  \caption{\label{tab:reg_results_mu} Posterior average marginal effects for overall WBI in Massachusetts (2021–2023) in the mean model. Estimates are reported on the original WBI scale (0–100), with 95\% highest posterior density credible intervals in parentheses. Bolded indicates statistical significance at the 5\% level.}
  \resizebox{\textwidth}{!}{%
  \begin{tabular}{@{\extracolsep{-1pt}}lrrr}
\\ [-1.8ex] \hline \hline \\
[-1.8ex] & \textit{Overall WBI (2021)} & \textit{Overall WBI (2022)} & \textit{Overall WBI (2023)} \\
\hline \\[-1.8ex]
\multicolumn{1}{l}{Sex (ref: Male)}\\
 ~~~~~~~~Female & \textbf{-1.1} (-2.2, -0.3) & \textbf{-0.9} (-1.9, 0) & 0.2 (-0.8, 1.3)\\ 
\multicolumn{1}{l}{Age} & \textbf{1.5} (0.9, 2.0) & \textbf{1.3} (0.7, 1.9) & 0.6 (-0.1, 1.2)\\
\multicolumn{1}{l}{Race (ref: White)}\\
  ~~~~~~~~Black & 0.5 (-1.5, 2.5) & 0.8 (-1.1, 2.6) & -1.1 (-2.5, 0.8)\\ 
  ~~~~~~~~Hispanic/Latino & \textbf{3.0} (1.3, 5.1) & 1.4 (-0.4, 2.9) & 1.8 (0, 3.3) \\ 
  ~~~~~~~~Asian & \textbf{4.3} (2.1, 6.3) & \textbf{2.1} (0.1, 4.1) & \textbf{3.3} (1.1, 5.1) \\ 
  ~~~~~~~~Other Race & -0.2 (-3.8, 2.9) & 2.1 (-1.9, 5.3) & 0 (-3.3, 4.2) \\ 
\multicolumn{1}{l}{Marital (ref: Never Married)}\\
  ~~~~~~~~Married & \textbf{1.6} (0.3, 2.7) & \textbf{2.5} (1.3, 3.8) & \textbf{2.7} (1.5, 4.1) \\ 
  ~~~~~~~~Other Marital Status & -1.0 (-2.8, 0.5) & -1.1 (-2.8, 0.7) & 0.5 (-1.4, 2.2) \\ 
  \multicolumn{1}{l}{Education (ref: $<$ High School)}\\
  ~~~~~~~~High School & -0.4 (-1.5, 0.8) & \textbf{-1.5} (-2.7, -0.2) & \textbf{-1.9} (-3.3, -0.6) \\ 
  ~~~~~~~~College & \textbf{1.0} (0.3, 1.7) & 0.4 (-0.2, 1.1) & -0.4 (-1.2, 0.5) \\ 
  ~~~~~~~~Post-Graduate & \textbf{3.7} (2.6, 4.8) & \textbf{2.6} (1.6, 3.7) & \textbf{2.1} (0.7, 3.2) \\ 
\multicolumn{1}{l}{Income (ref: $<$ 25K)}\\
  ~~~~~~~~Income: 25-50K & \textbf{2.6} (1.0, 4.4) & \textbf{3.5} (1.8, 5.6) & \textbf{4.4} (2.5, 6.5) \\ 
  ~~~~~~~~Income: 50-100K & \textbf{6.7} (5.2, 8.5) & \textbf{7.7} (6.0, 9.5) & \textbf{8.2} (6.5, 10.2) \\ 
  ~~~~~~~~Income: 100K+ & \textbf{13.2} (11.5, 14.9) & \textbf{13.4} (11.5, 15.1) & \textbf{13.4} (11.6, 15.5) \\ 
\hline
\hline \\[-1.8ex]
\end{tabular}
}
\end{table}

\subsubsection{Effects of individual-level covariates in the mean model} Across all three years, income emerged as the most influential predictor of WBI, with large, positive, and statistically significant effects (Table \ref{tab:reg_results_mu}). In each year, we observe that the average marginal effect for having an income greater than \$100,000 is about 13.4. That is, we expect individuals with annual incomes greater than \$100,000 to have, on average, overall WBI scores 13.4 points higher than those earning less than \$25,000, holding other covariates constant. The 95\% credible interval (CI) for this effect was $(11.5, 15.1)$ in 2022. Note that, all interpretations related to individual-level covariates are on the original WBI scale from 0 to 100 (Section \ref{subsubsec:marginal}). We also find strong, positive, and statistically significant effects for post-graduate education, with a average marginal effects ranging from 2.1 to 3.7. We see no statistically significant effect of having a college degree on overall WBI in 2022 and 2023, and only a modest average marginal effect of in 2021 (1.0; 95\% CI: 0.3, 1.7). In 2021 and 2022, the marginal effects of age are statistically significant at 1.5 and 1.3, respectively. Thus, on average, we expect a one standard deviation increase in age to lead to an about 1.5 unit increase in overall WBI, holding all else constant in 2021. The average marginal effect for married individuals (reference: never married) was also positive and statistically significant, with effect sizes increasing each year. In 2023, we estimate that, on average, a married individual has a overall WBI score 2.7 points higher than that of an individual who has never been married, holding all other variables constant. Notably, Asian individuals had higher WBI scores on average compared to White individuals, suggesting persistent racial heterogeneity in well-being outcomes. For example, in 2021, the average marginal effect for Asian was 4.3 (95\% CI: 2.1, 6.3). That is, on average we expect an Asian individual in 2021 to have an overall WBI 4.2 points higher than a White individual, holding all else constant. 

\begin{figure}[!]
\centering
\begin{tabular}{|c|c|c|}
\hline
\subfigure[2021 ZCTA Average Marginal Effects]{\label{fig:alphaMap21}\includegraphics[trim=0cm 0cm 0cm 0cm, clip, scale=0.42, page=1]{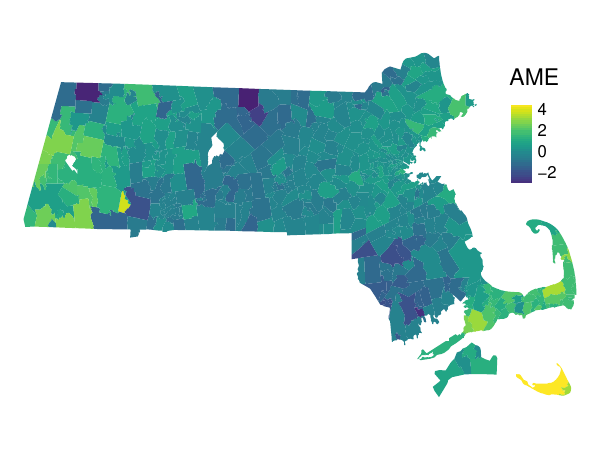}} & 
\subfigure[2022 ZCTA Average Marginal Effects]{\label{fig:alphaMap22}\includegraphics[trim=0cm 0cm 0cm 0cm, clip, scale=0.42, page=1]{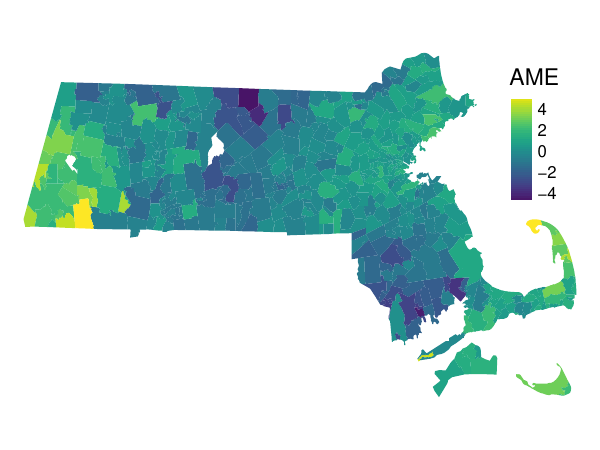}} &
\subfigure[2023 ZCTA Average Marginal Effects]{\label{fig:alphaMap23}\includegraphics[trim=0cm 0cm 0cm 0cm, clip, scale=0.42, page=1]{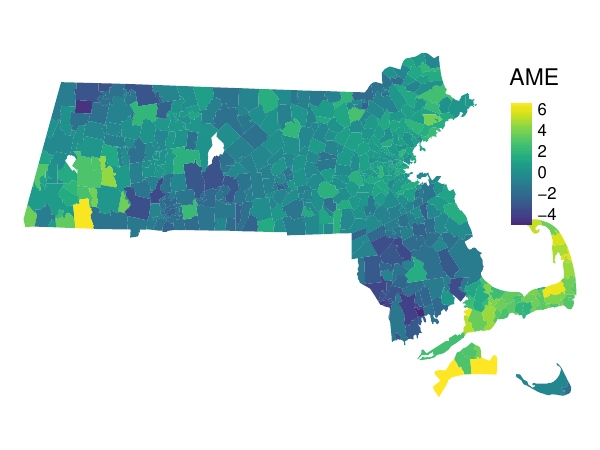}} \\ \hline
\subfigure[2021 ZCTA Quintiles]{\label{fig:rankMap21}\includegraphics[trim=0cm 0cm 0cm 0cm, clip, scale=0.42, page=2]{plots_model_zcta-choropleths2021.pdf}} &
\subfigure[2022 ZCTA Quintiles]{\label{fig:rankMap22}\includegraphics[trim=0cm 0cm 0cm 0cm, clip, scale=0.42, page=2]{plots_model_zcta-choropleths2022.pdf}} & 
\subfigure[2023 ZCTA Quintiles]{\label{fig:rankMap23}\includegraphics[trim=0cm 0cm 0cm 0cm, clip, scale=0.42, page=2]{plots_model_zcta-choropleths2023.pdf}} \\ \hline
\end{tabular}
\caption{Choropleth maps of ZCTA average marginal effects and corresponding quintiles for the years 2021–2023. Each map displays the spatial distribution of ZCTA-level effects on the WBI, adjusted for individual- and ZCTA-level covariates. The top panel in each year shows the estimated ZCTA effects, while the bottom panel presents the ZCTAs categorized by quintiles.}
\end{figure}

\subsubsection{ZCTA-level spatial effect estimates} ZCTA-level average marginal effects were relatively stable across years, with higher values concentrated in Western Massachusetts, Boston suburbs, and Cape Cod (Figures \ref{fig:alphaMap21}--\ref{fig:rankMap23}). For example, ZCTA 01255 (Sandisfield, in southwestern Massachusetts) was consistently one of the best performing ZCTAs, with effects of 2.7, 5.3, and 6.4 in 2021, 2022, and 2023, respectively. That is, in 2021, on average we would expect the WBI to increase by 2.7 points if they resided in Sandisfield as opposed to the average Massachusetts ZCTA, holding all else constant. In contrast, the ZCTAs located near and just west of Worcester, as well as those in southeastern Massachusetts, were identified as the lowest-performing areas. For example, ZCTA 02727 (North Dartmouth, in southeastern Massachusetts) had an average marginal effect of -2.3 in 2021, -3.1 in 2022, and -4.1 in 2023. Thus, in 2023, on average, we would expect an individual's WBI to decrease by 4.1 units if they lived in North Dartmouth compared to the average Massachusetts ZCTA.

\subsubsection{Variability in ZCTA-level effects over time} The distribution of ZCTA effects became more dispersed over time.  In 2021, the estimated ZCTA average marginal effects were concentrated around zero with 73.5\% of ZCTAs having effect sizes between -1 and 1. In 2022 and 2023, those percentages were 62.6\% and 49.7\%. This pattern reflects the influence of the prior structure: the mean-zero prior in 2021 shrank estimates toward zero, while informative priors in later years allowed for greater deviation based on prior information. Correspondingly, the range of effects widened from (-3.6, 4.3) in 2021 to (-4.7, 5.3) in 2022 and (-4.9, 6.6) in 2023. These results suggest that the Bayesian beta regression framework yields stable and interpretable ZCTA-level estimates (compared to \citet{jain2024modeling}, which used linear regression with spatial smoothing). Our approach produces more conservative and robust estimates, with fewer extreme outliers.

\subsubsection{Effects of ZCTA-level covariates in the precision model} None of the variables were statistically significant in the initial year 2021, but in 2022, both absolute age diversity and racial diversity indices were statistically significant, with positive coefficients (Table \ref{tab:reg_results_phi}). This means that as the difference in proportion between minors and seniors increases, we expect there to be greater intra-ZCTA precision or less intra-ZCTA variance in the WBI scores. Similarly, the racial diversity index coefficient is statistically significant and positive, indicating that as ZCTAs get more diverse, we should expect to see less intra-ZCTA variance in the WBI, holding all else constant. In contrast, we found no statistically significant effect for economic or educational segregation indices in any year. Note that the intercept of the precision model ranged from 1.52 and 2.25 across years, aligning with low to medium precision levels used in our simulation study (Section \ref{sec:simulation}). This supports the use of informative priors in later years, as they potentially enhance the estimation of ZCTA-level effects.

\begin{table}[!t]
\centering
  \caption{\label{tab:reg_results_phi} Posterior modes for coefficients in the precision model for overall WBI in Massachusetts (2021-2023). Bolded estimates denote statistical significance at the 5\% level and 95\% highest posterior density credible intervals are shown in parentheses.}
  \resizebox{\textwidth}{!}{%
  \begin{tabular}{@{\extracolsep{-1pt}}lrrr}
\hline
 & \textit{Overall WBI (2021)} & \textit{Overall WBI (2022)} & \textit{Overall WBI (2023)} \\
\hline
Intercept & \textbf{2.25} (1.66, 2.73) & \textbf{1.56} (0.74, 2.35) & \textbf{1.52} (0.87, 2.19) \\ 
\% Female & -0.27 (-1.46, 0.69) & 0.62 (-0.90, 2.17) & 0.69 (-0.65, 1.96)\\
Absolute Age diversity & 0.26 (-0.02, 0.59) & \textbf{0.39} (0.06, 0.75) & 0.28 (-0.08, 0.65)\\
Absolute economic segregation & 0.04 (-0.22, 0.29) & 0.18 (-0.07, 0.45) & -0.03 (-0.30, 0.26)\\
Absolute educational segregation & 0.02 (-0.22, 0.27) & -0.07 (-0.28, 0.19) & 0.25 (-0.03, 0.48)\\
Racial diversity index & 0.04 (-0.22, 0.29) & \textbf{0.40} (0.16, 0.69) & 0.28 (-0.05, 0.55)\\
\hline \\[-1.8ex]
\end{tabular}
}
\end{table}

\subsection{Computational performance}\label{subsec:time}
To assess the computational demands of our Bayesian beta regression model, we recorded the runtime for the MCMC sampling across all three years of analysis. Model estimation was performed using Stan, with each chain executed on a separate node of our institution's computing cluster via batch job submission. Each chain consisted of 4,000 warm-up iterations followed by 6,000 post-warm-up samples. In each year, the model specification included 14 covariates in the mean model and 6 covariates in the precision model. In 2021, the dataset comprised 4,443 individual-level observations. The average warm-up and sampling times per chain were 12.8 and 19.8 hours, respectively, yielding a total average runtime of 32.6 hours. For 2022, with 3,602 observations, the average warm-up and sampling times were 14.6 and 31.1 hours, respectively, resulting in a total runtime of 45.7 hours. In 2023, there were 3,428 observations and the average warm-up and sampling times were 10.8 hours and 16.5 hours, respectively, for a total average runtime of 27.3 hours. These results provide an estimate for the computational time of fully Bayesian estimation for our high-dimensional hierarchical modeling framework, particularly when incorporating spatially- and temporally-informed priors and modeling both mean and precision components.

\section{Discussion}\label{sec:discussion}
In this paper, we develop a Bayesian double-generalized beta regression framework with spatial and temporal borrowing to model individual well-being across Massachusetts from 2021 to 2023. Our novel approach incorporates individual-level covariates in the mean model and ZCTA-level covariates in the precision model, while leveraging spatial structure and prior information across time to improve estimation, particularly in data-sparse regions. The use of ZCTA-level covariates in the precision model is especially novel in this context as variability in well-being is expected to be impacted by various ZCTA-level factors. Through an extensive, realistic simulation study, we showed that the model is able to accurately estimate the effects of model covariates and ZCTA-level spatial effects, while capturing the spatial distribution of ZCTA-level effects. Simulation results further demonstrated that Bayesian borrowing in the form of incorporating posterior estimates from one year as priors in subsequent years enhances the estimation of ZCTA-level effects, particularly in regions with limited data. However, we did not however compare our model performance to other methods, as the bounded nature of our outcome modeling results in incongruous comparisons to methods with unbounded support and also since there is a lack of well-established methods for well-being modeling. This method has broad applications and can be used to model both individual- and population-level health outcomes and indices that are spatial in nature and measured over time. We have also developed an R package called \texttt{BayesBadger} that provides software to implement a Bayesian double generalized beta regression model with individual-level data used to model the mean and cluster-level data to model the precision. The model can also incorporate spatial smoothing through a graph Laplacian prior and includes functions to compute marginal effects of the mean model covariates as well as the cluster or spatial-level effects. The \texttt{BayesBadger} R package can be found at \url{https://github.com/abhijainstats/BayesBadger}.

In our case study of Massachusetts, our findings highlight income, education, and marital status as the most influential individual-level predictors of well-being, each exhibiting consistently positive effects across years. At the geographic level, ZCTA spatial effects revealed persistent spatial patterns: areas in western Massachusetts, Cape Cod, and the Boston suburbs consistently performed among the best in terms of well-being, while ZCTAs near Worcester and in southeastern Massachusetts had the lowest estimates. These spatial disparities remained consistent over time. Overall, our dual-borrowing approach contributes to the stability observed in the ZCTA-level spatial effects over time. We also employed driving times between ZCTA population centroids to inform our adjacency matrix, which gives a more realistic measure of accessibility to neighboring ZCTAs' resources and amenities.

Despite these strengths, certain limitations warrant discussion. First, the dataset exhibits substantial geographic sparsity: each year, approximately 90-100 ZCTAs were entirely unrepresented, and many others had only a handful of observations. While the use of a graph Laplacian based prior enables partial information sharing across neighboring ZCTAs, a more geographically balanced sample would improve the estimation accuracy of both individual- and ZCTA-level estimates. Second, we use only three years worth of survey data, which limits the strength of temporal borrowing and doesn't allows for the temporal dynamics to be fully captured. With additional years of data, we could see potentially more meaningful differences in ZCTA-level effects as the built environment and population dynamics of a ZCTA change; a three-year period is likely not enough time to realize those changes. Third, the sample may not be representative of the overall Massachusetts population. For example, the 2021 sample was over 82\% White, decreasing to 66\% by 2023. At least 72\% of the sample respondents held at least a Bachelor's degree. These demographic skews may limit the generalizability of our findings. Fourth, the computational burden of the model is substantial: each model requiring 24-48 hours to complete, hindering its implementation for a nation-wide analysis. 

Future work will focus on improving computational efficiency to enhance scalability. One promising direction is the use of approximation methods such as Integrated Nested Laplace Approximation (INLA), which may substantially reduce runtime while preserving inference quality \citep{lindgren2015bayesian}. Such improvements would also enable the development of a unified spatiotemporal model that jointly estimates well-being across all years, rather than fitting separate models for each year. Other possible extensions include models that can account for excess zeros or ones, by adding in a zero-or-one-inflated component to the model.

\bibliographystyle{elsarticle-harv} 
\bibliography{bib}

\clearpage
\section*{Appendix}
\appendix
\section{Derivation of posterior distribution and full conditional of all model parameters}

Here is a table with definitions and priors for all model parameters:
\begin{table}[ht]
\centering
\caption{Model parameters and prior distributions.}
\label{tab:priors}
\begin{tabular}{>{\centering}m{2.5cm} >{\raggedright}m{6cm} >{\raggedright\arraybackslash}m{5cm}}
\toprule
\textbf{Parameter} & \textbf{Definition} & \textbf{Prior} \\
\midrule
$\boldsymbol{\alpha}_t$ 
  & $S$-dimensional vector of ZCTA-level spatial effects for year $t$ 
  & $\begin{cases} 
      N_S(\mathbf{0},\, M^{-1}) & \text{if } t = 1 \\ 
      N_S(\widehat{\boldsymbol{\alpha}}_{t-1},\, M^{-1}) & \text{if } t > 1 
    \end{cases}$ \\[8pt]
\hline
$\boldsymbol{\beta}_t$ 
    & $p$-dimensional vector of regression coefficients for the mean model
    & $N_p(\mathbf{0}, \tau_{\mu}I_p)$ \\[8pt]
\hline
$\boldsymbol{\omega}_t$ 
    & $q$-dimensional vector of regression coefficients for the dispersion model
    & $N_q(\mathbf{0}, \tau_{\phi}I_q)$ \\[8pt]
\hline
$\lambda$ 
    & Scalar spatial smoothing parameter
    & $\text{Gamma}(\theta_1, \theta_2)$ \\[8pt]
\hline
$\gamma$ 
    & Scalar numerical stability parameter
    & $\text{Gamma}(\iota_1, \iota_2)$ \\[8pt]
\bottomrule
\end{tabular}
\end{table}

The joint posterior distribution for the initial year is then:
\begin{align*}
    \pi(\boldsymbol{\beta}, \boldsymbol{\alpha}, \boldsymbol{\omega}, \lambda, \gamma | y) & \propto f(y|\boldsymbol{\beta}, \alpha, \boldsymbol{\omega}, \lambda, \gamma) \pi(\boldsymbol{\beta}) \pi_1(\boldsymbol{\alpha}|\lambda, \gamma) \pi(\boldsymbol{\omega}) \pi(\lambda) \pi(\gamma) \\
    & \propto \prod_{s=1}^{S} \prod_{s: i=1}^{n_s} \frac{\Gamma(\phi_{s})}{\Gamma(\mu_{is}\phi_{s}) \Gamma((1-\mu_{is})\phi_{s})} 
    y^{\mu_{is} \phi_{s} -1} (1-y)^{(1-\mu_{is}) \phi_{s} -1} \\
    & \quad \times \pi(\boldsymbol{\beta}) \pi_1(\boldsymbol{\alpha}) \pi(\boldsymbol{\omega}) \pi(\lambda) \pi(\gamma) \\
    & \propto \Bigg(\prod_{s=1}^{S} \prod_{s: i=1}^{n_s} \frac{\Gamma(\phi_{s})}{\Gamma(\mu_{is}\phi_{s}) \Gamma((1-\mu_{is})\phi_{s})}\Bigg) \\
    & \quad \times y^{\left(\sum_{s=1}^S \sum_{s:i=1}^{n_s} \mu_{is} \phi_{s}\right)-N} 
    (1-y)^{\left(\sum_{s=1}^S \sum_{s:i=1}^{n_s} (1-\mu_{is}) \phi_{s}\right)-N} \\
    & \quad \times \pi(\boldsymbol{\beta}) \pi_1(\boldsymbol{\alpha}) \pi(\boldsymbol{\omega}) \pi(\lambda) \pi(\gamma) \\
    & \propto \Bigg(\prod_{s=1}^{S} \prod_{s: i=1}^{n_s} \frac{\Gamma\left(e^{c_{s}\omega_{s}}\right)}    {\Gamma \left(\frac{e^{\mathbf{x}_{is}\boldsymbol{\beta}+\boldsymbol{\alpha}_s+\mathbf{c}_s \boldsymbol{\omega}_s}}{1+e^{\mathbf{x}_{is}\boldsymbol{\beta}+\boldsymbol{\alpha}_s}}\right)
    \Gamma\left(e^{\mathbf{c}_s \boldsymbol{\omega}_s}- \frac{e^{\mathbf{x}_{is}\boldsymbol{\beta}+\boldsymbol{\alpha}_s + \mathbf{c}_s\boldsymbol{\omega}_s}}{1+e^{\mathbf{x}_{is}\boldsymbol{\beta}+\boldsymbol{\alpha}_s}}\right)}\Bigg) \\
    & \quad \times y^{\left(\sum_{s=1}^S \sum_{s:i=1}^{n_s} \frac{e^{\mathbf{x}_{is}\boldsymbol{\beta}+\alpha_s + \mathbf{c}_s \boldsymbol{\omega}_s}}{1+e^{\mathbf{x}_{is}\boldsymbol{\beta}+\alpha_s}}\right)-N} \\
    & \quad \times (1-y)^{\left(\sum_{s=1}^S \sum_{s:i=1}^{n_s} e^{\mathbf{c}_s \boldsymbol{\omega}_s} - \frac{e^{\mathbf{x}_{is}\boldsymbol{\beta}+\alpha_s + \mathbf{c}_s \boldsymbol{\omega}_s}}{1+e^{\mathbf{x}_{is}\boldsymbol{\beta}+\alpha_s }}\right)-N} \\
    & \quad \times \Big((2\pi)^{-\frac{p}{2}} |\tau_{\mu}I_p|^{-\frac{1}{2}} e^{-\frac{1}{2} (\boldsymbol{\beta} - \mathbf{0})^T (\tau_{\mu}I_p)^{-1} (\boldsymbol{\beta} - \mathbf{0})} \Big) \\
    & \quad \times \Big((2\pi)^{-\frac{S}{2}} |(\lambda L + \lambda\gamma I_S)^{-1}|^{-\frac{1}{2}} e^{-\frac{1}{2} (\boldsymbol{\alpha} - \mathbf{0})^T (\lambda L + \lambda\gamma I_S) (\boldsymbol{\alpha - \mathbf{0})}} \Big) \\
    & \quad \times \Big((2\pi)^{-\frac{q}{2}} |\tau_{\phi}I_q|^{-\frac{1}{2}} e^{-\frac{1}{2} (\boldsymbol{\omega} - \mathbf{0})^T (\tau_{\phi}I_q)^{-1} (\boldsymbol{\omega} - \mathbf{0})} \Big) \\
    & \quad \times \frac{\theta_2 \theta_1}{\Gamma(\theta_1)} \lambda^{\theta_1-1} e^{-\theta_2 \lambda} \frac{\iota_2 \iota_1}{\Gamma(\iota_1)} \gamma^{\iota_1-1} e^{-\iota_2 \gamma}
\end{align*}

The full conditional for $\boldsymbol{\alpha}$ can be expressed as:
\begin{align*}
    \pi(\boldsymbol{\alpha}|\boldsymbol{\beta},\boldsymbol{\omega}, \lambda, \gamma, y) & \propto \Bigg(\prod_{s=1}^{S} \prod_{s: i=1}^{n_s} \frac{1} {\Gamma \left(\frac{e^{\mathbf{x}_{is}\boldsymbol{\beta}+\boldsymbol{\alpha}_s+\mathbf{c}_s \boldsymbol{\omega}_s}}{1+e^{\mathbf{x}_{is}\boldsymbol{\beta}+\boldsymbol{\alpha}_s}}\right)
    \Gamma\left(e^{\mathbf{c}_s \boldsymbol{\omega}_s}- \frac{e^{\mathbf{x}_{is}\boldsymbol{\beta}+\boldsymbol{\alpha}_s + \mathbf{c}_s\boldsymbol{\omega}_s}}{1+e^{\mathbf{x}_{is}\boldsymbol{\beta}+\boldsymbol{\alpha}_s}}\right)}\Bigg) \\
    & \quad \times y^{\left(\sum_{s=1}^S \sum_{s:i=1}^{n_s} \frac{e^{\mathbf{x}_{is}\boldsymbol{\beta}+\alpha_s + \mathbf{c}_s \boldsymbol{\omega}_s}}{1+e^{\mathbf{x}_{is}\boldsymbol{\beta}+\alpha_s}}\right)} (1-y)^{\left(\sum_{s=1}^S \sum_{s:i=1}^{n_s} e^{\mathbf{c}_s \boldsymbol{\omega}_s} - \frac{e^{\mathbf{x}_{is}\boldsymbol{\beta}+\alpha_s + \mathbf{c}_s \boldsymbol{\omega}_s}}{1+e^{\mathbf{x}_{is}\boldsymbol{\beta}+\alpha_s }}\right)}\\
    & \quad \times \Big(e^{-\frac{1}{2} (\boldsymbol{\alpha} - \mathbf{0})^T (\lambda L + \lambda\gamma I_S) (\boldsymbol{\alpha - \mathbf{0})}} \Big)
\end{align*}

The full conditional for $\boldsymbol{\beta}$ can be expressed as:
\begin{align*}
    \pi(\boldsymbol{\beta}|\boldsymbol{\alpha},\boldsymbol{\omega}, \lambda, \gamma, y) & \propto \Bigg(\prod_{s=1}^{S} \prod_{s: i=1}^{n_s} \frac{1} {\Gamma \left(\frac{e^{\mathbf{x}_{is}\boldsymbol{\beta}+\boldsymbol{\alpha}_s+\mathbf{c}_s \boldsymbol{\omega}_s}}{1+e^{\mathbf{x}_{is}\boldsymbol{\beta}+\boldsymbol{\alpha}_s}}\right)
    \Gamma\left(e^{\mathbf{c}_s \boldsymbol{\omega}_s}- \frac{e^{\mathbf{x}_{is}\boldsymbol{\beta}+\boldsymbol{\alpha}_s + \mathbf{c}_s\boldsymbol{\omega}_s}}{1+e^{\mathbf{x}_{is}\boldsymbol{\beta}+\boldsymbol{\alpha}_s}}\right)}\Bigg) \\
    & \quad \times y^{\left(\sum_{s=1}^S \sum_{s:i=1}^{n_s} \frac{e^{\mathbf{x}_{is}\boldsymbol{\beta}+\alpha_s + \mathbf{c}_s \boldsymbol{\omega}_s}}{1+e^{\mathbf{x}_{is}\boldsymbol{\beta}+\alpha_s}}\right)} (1-y)^{\left(\sum_{s=1}^S \sum_{s:i=1}^{n_s} e^{\mathbf{c}_s \boldsymbol{\omega}_s} - \frac{e^{\mathbf{x}_{is}\boldsymbol{\beta}+\alpha_s + \mathbf{c}_s \boldsymbol{\omega}_s}}{1+e^{\mathbf{x}_{is}\boldsymbol{\beta}+\alpha_s }}\right)}\\
    & \quad \times \Big(e^{-\frac{1}{2} (\boldsymbol{\beta} - \mathbf{0})^T (\tau_{\mu}I_p)^{-1} (\boldsymbol{\beta} - \mathbf{0})} \Big)
\end{align*}

The full conditional for $\boldsymbol{\omega}$ can be expressed as:
\begin{align*}
    \pi(\boldsymbol{\omega}|\boldsymbol{\alpha},\boldsymbol{\beta}, \lambda, \gamma, y) & \propto \Bigg(\prod_{s=1}^{S} \prod_{s: i=1}^{n_s} \frac{1} {\Gamma \left(\frac{e^{\mathbf{x}_{is}\boldsymbol{\beta}+\boldsymbol{\alpha}_s+\mathbf{c}_s \boldsymbol{\omega}_s}}{1+e^{\mathbf{x}_{is}\boldsymbol{\beta}+\boldsymbol{\alpha}_s}}\right)
    \Gamma\left(e^{\mathbf{c}_s \boldsymbol{\omega}_s}- \frac{e^{\mathbf{x}_{is}\boldsymbol{\beta}+\boldsymbol{\alpha}_s + \mathbf{c}_s\boldsymbol{\omega}_s}}{1+e^{\mathbf{x}_{is}\boldsymbol{\beta}+\boldsymbol{\alpha}_s}}\right)}\Bigg) \\
    & \quad \times y^{\left(\sum_{s=1}^S \sum_{s:i=1}^{n_s} \frac{e^{\mathbf{x}_{is}\boldsymbol{\beta}+\alpha_s + \mathbf{c}_s \boldsymbol{\omega}_s}}{1+e^{\mathbf{x}_{is}\boldsymbol{\beta}+\alpha_s}}\right)} (1-y)^{\left(\sum_{s=1}^S \sum_{s:i=1}^{n_s} e^{\mathbf{c}_s \boldsymbol{\omega}_s} - \frac{e^{\mathbf{x}_{is}\boldsymbol{\beta}+\alpha_s + \mathbf{c}_s \boldsymbol{\omega}_s}}{1+e^{\mathbf{x}_{is}\boldsymbol{\beta}+\alpha_s }}\right)}\\
    & \quad \times \Big(e^{-\frac{1}{2} (\boldsymbol{\omega} - \mathbf{0})^T (\tau_{\phi}I_q)^{-1} (\boldsymbol{\omega} - \mathbf{0})} \Big)
\end{align*}

The full conditional for $\boldsymbol{\lambda}$ can be expressed as:
\begin{align*}
    \pi(\lambda|\boldsymbol{\beta},\boldsymbol{\alpha},\boldsymbol{\omega}, \gamma, y) & \propto
    \Big(|(\lambda L + \lambda\gamma I_S)^{-1}|^{-\frac{1}{2}} e^{-\frac{1}{2} (\boldsymbol{\alpha} - \mathbf{0})^T (\lambda L + \lambda\gamma I_S) (\boldsymbol{\alpha - \mathbf{0})}} \Big) \\
    & \quad \times \lambda^{\theta_1-1} e^{-\theta_2 \lambda}
\end{align*}

The full conditional for $\boldsymbol{\gamma}$ can be expressed as:
\begin{align*}
    \pi(\gamma|\boldsymbol{\beta},\boldsymbol{\alpha},\boldsymbol{\omega}, \lambda, y) & \propto
    \Big(|(\lambda L + \lambda\gamma I_S)^{-1}|^{-\frac{1}{2}} e^{-\frac{1}{2} (\boldsymbol{\alpha} - \mathbf{0})^T (\lambda L + \lambda\gamma I_S) (\boldsymbol{\alpha - \mathbf{0})}} \Big) \\
    & \quad \times \gamma^{\iota_1-1} e^{-\iota_2 \gamma}
\end{align*}

\section{Additional tables and figures for the Simulation study}

\begin{table}[H]
\centering
\caption{\label{tab:sim_coef_smooth}95\% credible interval coverage (in percent) of mean and precision model parameters for the Smooth pattern across precision settings, years, and prior specifications. ``NI'' = non-informative prior; ``I'' = informative prior using previous year's posterior estimates.}
\small
\resizebox{\textwidth}{!}{%
\begin{tabular}{c|ccccc|ccccc|ccccc|}
\cline{2-16}
& \multicolumn{5}{c|}{Low Precision} & \multicolumn{5}{c|}{Medium Precision} & \multicolumn{5}{c|}{High Precision} \\
\cline{2-16}
& \multicolumn{1}{c|}{2021} & \multicolumn{2}{c|}{2022} & \multicolumn{2}{c|}{2023}
& \multicolumn{1}{c|}{2021} & \multicolumn{2}{c|}{2022} & \multicolumn{2}{c|}{2023}
& \multicolumn{1}{c|}{2021} & \multicolumn{2}{c|}{2022} & \multicolumn{2}{c|}{2023} \\
\hline
\multicolumn{1}{|c|}{Parameter} & \multicolumn{1}{c|}{NI} & NI & \multicolumn{1}{c|}{I} & NI & I
& \multicolumn{1}{c|}{NI} & NI & \multicolumn{1}{c|}{I} & NI & I
& \multicolumn{1}{c|}{NI} & NI & \multicolumn{1}{c|}{I} & NI & I \\ \hline
\multicolumn{1}{|c|}{$\beta_1$} & \multicolumn{1}{c|}{93} & 90 & \multicolumn{1}{c|}{90} & 83 & 87 & \multicolumn{1}{c|}{90} & 80 & \multicolumn{1}{c|}{83} & 100 & 100 & \multicolumn{1}{c|}{97} & 90 & \multicolumn{1}{c|}{100} & 100 & 100 \\
\multicolumn{1}{|c|}{$\beta_2$} & \multicolumn{1}{c|}{90} & 83 & \multicolumn{1}{c|}{87} & 93 & 90 & \multicolumn{1}{c|}{90} & 100 & \multicolumn{1}{c|}{100} & 100 & 100 & \multicolumn{1}{c|}{100} & 93 & \multicolumn{1}{c|}{93} & 97 & 97 \\
\multicolumn{1}{|c|}{$\beta_3$} & \multicolumn{1}{c|}{93} & 90 & \multicolumn{1}{c|}{93} & 93 & 93 & \multicolumn{1}{c|}{93} & 100 & \multicolumn{1}{c|}{93} & 100 & 97 & \multicolumn{1}{c|}{93} & 93 & \multicolumn{1}{c|}{90} & 97 & 97 \\
\multicolumn{1}{|c|}{$\beta_4$} & \multicolumn{1}{c|}{90} & 90 & \multicolumn{1}{c|}{77} & 87 & 80 & \multicolumn{1}{c|}{93} & 97 & \multicolumn{1}{c|}{93} & 100 & 100 & \multicolumn{1}{c|}{93} & 80 & \multicolumn{1}{c|}{80} & 93 & 97 \\
\multicolumn{1}{|c|}{$\beta_5$} & \multicolumn{1}{c|}{90} & 90 & \multicolumn{1}{c|}{83} & 80 & 73 & \multicolumn{1}{c|}{97} & 93 & \multicolumn{1}{c|}{87} & 100 & 97 & \multicolumn{1}{c|}{87} & 93 & \multicolumn{1}{c|}{93} & 97 & 97 \\ \hline
\multicolumn{1}{|c|}{$\omega_0$} & \multicolumn{1}{c|}{83} & 97 & \multicolumn{1}{c|}{97} & 93 & 93 & \multicolumn{1}{c|}{90} & 93 & \multicolumn{1}{c|}{93} & 100 & 100 & \multicolumn{1}{c|}{70} & 93 & \multicolumn{1}{c|}{100} & 93 & 93 \\
\multicolumn{1}{|c|}{$\omega_1$} & \multicolumn{1}{c|}{70} & 87 & \multicolumn{1}{c|}{87} & 90 & 93 & \multicolumn{1}{c|}{77} & 90 & \multicolumn{1}{c|}{93} & 93 & 93 & \multicolumn{1}{c|}{83} & 87 & \multicolumn{1}{c|}{87} & 90 & 90 \\
\multicolumn{1}{|c|}{$\omega_2$} & \multicolumn{1}{c|}{83} & 90 & \multicolumn{1}{c|}{90} & 93 & 97 & \multicolumn{1}{c|}{93} & 97 & \multicolumn{1}{c|}{93} & 93 & 90 & \multicolumn{1}{c|}{93} & 87 & \multicolumn{1}{c|}{90} & 93 & 93 \\
\multicolumn{1}{|c|}{$\omega_3$} & \multicolumn{1}{c|}{80} & 93 & \multicolumn{1}{c|}{90} & 100 & 100 & \multicolumn{1}{c|}{93} & 97 & \multicolumn{1}{c|}{93} & 90 & 90 & \multicolumn{1}{c|}{90} & 97 & \multicolumn{1}{c|}{97} & 90 & 87 \\
\hline
\end{tabular}
}
\end{table}

\begin{table}[H]
\centering
\caption{\label{tab:sim_coef_hotspots}95\% credible interval coverage (in percent) of mean and precision model parameters for the Hotspots pattern across precision settings, years, and prior specifications. ``NI'' = non-informative prior; ``I'' = informative prior using previous year's posterior estimates.}
\small
\resizebox{\textwidth}{!}{%
\begin{tabular}{c|ccccc|ccccc|ccccc|}
\cline{2-16}
& \multicolumn{5}{c|}{Low Precision} & \multicolumn{5}{c|}{Medium Precision} & \multicolumn{5}{c|}{High Precision} \\
\cline{2-16}
& \multicolumn{1}{c|}{2021} & \multicolumn{2}{c|}{2022} & \multicolumn{2}{c|}{2023}
& \multicolumn{1}{c|}{2021} & \multicolumn{2}{c|}{2022} & \multicolumn{2}{c|}{2023}
& \multicolumn{1}{c|}{2021} & \multicolumn{2}{c|}{2022} & \multicolumn{2}{c|}{2023} \\
\hline
\multicolumn{1}{|c|}{Parameter} & \multicolumn{1}{c|}{NI} & NI & \multicolumn{1}{c|}{I} & NI & I
& \multicolumn{1}{c|}{NI} & NI & \multicolumn{1}{c|}{I} & NI & I
& \multicolumn{1}{c|}{NI} & NI & \multicolumn{1}{c|}{I} & NI & I \\ \hline
\multicolumn{1}{|c|}{$\beta_1$} & \multicolumn{1}{c|}{97} & 90 & \multicolumn{1}{c|}{90} & 100 & 100 & \multicolumn{1}{c|}{93} & 97 & \multicolumn{1}{c|}{97} & 97 & 100 & \multicolumn{1}{c|}{90} & 87 & \multicolumn{1}{c|}{87} & 97 & 97 \\
\multicolumn{1}{|c|}{$\beta_2$} & \multicolumn{1}{c|}{100} & 93 & \multicolumn{1}{c|}{93} & 93 & 90 & \multicolumn{1}{c|}{93} & 97 & \multicolumn{1}{c|}{97} & 100 & 97 & \multicolumn{1}{c|}{93} & 90 & \multicolumn{1}{c|}{97} & 93 & 93 \\
\multicolumn{1}{|c|}{$\beta_3$} & \multicolumn{1}{c|}{93} & 97 & \multicolumn{1}{c|}{97} & 90 & 90 & \multicolumn{1}{c|}{83} & 100 & \multicolumn{1}{c|}{100} & 93 & 93 & \multicolumn{1}{c|}{87} & 97 & \multicolumn{1}{c|}{90} & 97 & 97 \\
\multicolumn{1}{|c|}{$\beta_4$} & \multicolumn{1}{c|}{97} & 93 & \multicolumn{1}{c|}{93} & 93 & 90 & \multicolumn{1}{c|}{87} & 93 & \multicolumn{1}{c|}{93} & 97 & 97 & \multicolumn{1}{c|}{97} & 93 & \multicolumn{1}{c|}{93} & 97 & 97 \\
\multicolumn{1}{|c|}{$\beta_5$} & \multicolumn{1}{c|}{90} & 83 & \multicolumn{1}{c|}{87} & 97 & 97 & \multicolumn{1}{c|}{90} & 97 & \multicolumn{1}{c|}{97} & 97 & 100 & \multicolumn{1}{c|}{90} & 93 & \multicolumn{1}{c|}{93} & 100 & 100 \\ \hline
\multicolumn{1}{|c|}{$\omega_0$} & \multicolumn{1}{c|}{93} & 100 & \multicolumn{1}{c|}{100} & 97 & 93 & \multicolumn{1}{c|}{87} & 90 & \multicolumn{1}{c|}{90} & 100 & 100 & \multicolumn{1}{c|}{90} & 87 & \multicolumn{1}{c|}{87} & 93 & 93 \\
\multicolumn{1}{|c|}{$\omega_1$} & \multicolumn{1}{c|}{73} & 93 & \multicolumn{1}{c|}{93} & 80 & 83 & \multicolumn{1}{c|}{77} & 90 & \multicolumn{1}{c|}{90} & 90 & 90 & \multicolumn{1}{c|}{73} & 93 & \multicolumn{1}{c|}{93} & 87 & 83 \\
\multicolumn{1}{|c|}{$\omega_2$} & \multicolumn{1}{c|}{87} & 93 & \multicolumn{1}{c|}{93} & 87 & 83 & \multicolumn{1}{c|}{90} & 87 & \multicolumn{1}{c|}{87} & 90 & 90 & \multicolumn{1}{c|}{93} & 87 & \multicolumn{1}{c|}{87} & 97 & 97 \\
\multicolumn{1}{|c|}{$\omega_3$} & \multicolumn{1}{c|}{90} & 93 & \multicolumn{1}{c|}{93} & 97 & 93 & \multicolumn{1}{c|}{97} & 90 & \multicolumn{1}{c|}{87} & 100 & 97 & \multicolumn{1}{c|}{87} & 93 & \multicolumn{1}{c|}{97} & 87 & 90 \\
\hline
\end{tabular}
}
\end{table}

\begin{table}[H]
\centering
\caption{\label{tab:sim_coef_random}95\% credible interval coverage (in percent) of mean and precision model parameters for the Random pattern across precision settings, years, and prior specifications. ``NI'' = non-informative prior; ``I'' = informative prior using previous year's posterior estimates.}
\small
\resizebox{\textwidth}{!}{%
\begin{tabular}{c|ccccc|ccccc|ccccc|}
\cline{2-16}
& \multicolumn{5}{c|}{Low Precision} & \multicolumn{5}{c|}{Medium Precision} & \multicolumn{5}{c|}{High Precision} \\
\cline{2-16}
& \multicolumn{1}{c|}{2021} & \multicolumn{2}{c|}{2022} & \multicolumn{2}{c|}{2023}
& \multicolumn{1}{c|}{2021} & \multicolumn{2}{c|}{2022} & \multicolumn{2}{c|}{2023}
& \multicolumn{1}{c|}{2021} & \multicolumn{2}{c|}{2022} & \multicolumn{2}{c|}{2023} \\
\hline
\multicolumn{1}{|c|}{Parameter} & \multicolumn{1}{c|}{NI} & NI & \multicolumn{1}{c|}{I} & NI & I
& \multicolumn{1}{c|}{NI} & NI & \multicolumn{1}{c|}{I} & NI & I
& \multicolumn{1}{c|}{NI} & NI & \multicolumn{1}{c|}{I} & NI & I \\ \hline
\multicolumn{1}{|c|}{$\beta_1$} & \multicolumn{1}{c|}{87} & 90 & \multicolumn{1}{c|}{93} & 93 & 97 & \multicolumn{1}{c|}{90} & 90 & \multicolumn{1}{c|}{93} & 90 & 90 & \multicolumn{1}{c|}{97} & 97 & \multicolumn{1}{c|}{97} & 97 & 97 \\
\multicolumn{1}{|c|}{$\beta_2$} & \multicolumn{1}{c|}{90} & 97 & \multicolumn{1}{c|}{97} & 87 & 87 & \multicolumn{1}{c|}{93} & 93 & \multicolumn{1}{c|}{93} & 100 & 97 & \multicolumn{1}{c|}{87} & 93 & \multicolumn{1}{c|}{93} & 97 & 97 \\
\multicolumn{1}{|c|}{$\beta_3$} & \multicolumn{1}{c|}{93} & 93 & \multicolumn{1}{c|}{97} & 83 & 80 & \multicolumn{1}{c|}{97} & 93 & \multicolumn{1}{c|}{93} & 93 & 90 & \multicolumn{1}{c|}{97} & 87 & \multicolumn{1}{c|}{83} & 93 & 90 \\
\multicolumn{1}{|c|}{$\beta_4$} & \multicolumn{1}{c|}{83} & 93 & \multicolumn{1}{c|}{93} & 90 & 87 & \multicolumn{1}{c|}{100} & 83 & \multicolumn{1}{c|}{87} & 93 & 90 & \multicolumn{1}{c|}{93} & 83 & \multicolumn{1}{c|}{87} & 90 & 93 \\
\multicolumn{1}{|c|}{$\beta_5$} & \multicolumn{1}{c|}{97} & 93 & \multicolumn{1}{c|}{93} & 80 & 83 & \multicolumn{1}{c|}{100} & 93 & \multicolumn{1}{c|}{93} & 90 & 87 & \multicolumn{1}{c|}{93} & 93 & \multicolumn{1}{c|}{93} & 93 & 90 \\ \hline
\multicolumn{1}{|c|}{$\omega_0$} & \multicolumn{1}{c|}{93} & 90 & \multicolumn{1}{c|}{90} & 87 & 87 & \multicolumn{1}{c|}{100} & 93 & \multicolumn{1}{c|}{93} & 93 & 90 & \multicolumn{1}{c|}{50} & 17 & \multicolumn{1}{c|}{20} & 50 & 60 \\
\multicolumn{1}{|c|}{$\omega_1$} & \multicolumn{1}{c|}{87} & 97 & \multicolumn{1}{c|}{97} & 90 & 93 & \multicolumn{1}{c|}{57} & 73 & \multicolumn{1}{c|}{73} & 87 & 90 & \multicolumn{1}{c|}{0} & 0 & \multicolumn{1}{c|}{0} & 50 & 43 \\
\multicolumn{1}{|c|}{$\omega_2$} & \multicolumn{1}{c|}{90} & 93 & \multicolumn{1}{c|}{93} & 97 & 97 & \multicolumn{1}{c|}{90} & 80 & \multicolumn{1}{c|}{80} & 93 & 93 & \multicolumn{1}{c|}{60} & 20 & \multicolumn{1}{c|}{17} & 47 & 53 \\
\multicolumn{1}{|c|}{$\omega_3$} & \multicolumn{1}{c|}{87} & 87 & \multicolumn{1}{c|}{87} & 90 & 90 & \multicolumn{1}{c|}{87} & 93 & \multicolumn{1}{c|}{90} & 97 & 93 & \multicolumn{1}{c|}{80} & 73 & \multicolumn{1}{c|}{73} & 83 & 87 \\
\hline
\end{tabular}
}
\end{table}

\begin{table}[H]
\centering
\caption{\label{tab:sim_rmse_fmt}RMSE of ZIP Code effects by spatial pattern across precision settings, years, and prior specifications. ``NI'' = non-informative prior; ``I'' = informative prior using previous year's posterior estimates.}
\small
\resizebox{\textwidth}{!}{%
\begin{tabular}{c|ccccc|ccccc|ccccc|}
\cline{2-16}
 & \multicolumn{5}{c|}{Low Precision} & \multicolumn{5}{c|}{Medium Precision} & \multicolumn{5}{c|}{High Precision} \\ \cline{2-16} 
 & \multicolumn{1}{c|}{2021} & \multicolumn{2}{c|}{2022} & \multicolumn{2}{c|}{2023} 
 & \multicolumn{1}{c|}{2021} & \multicolumn{2}{c|}{2022} & \multicolumn{2}{c|}{2023} 
 & \multicolumn{1}{c|}{2021} & \multicolumn{2}{c|}{2022} & \multicolumn{2}{c|}{2023} \\ \hline
\multicolumn{1}{|c|}{Pattern} & 
\multicolumn{1}{c|}{NI} & NI & \multicolumn{1}{c|}{I} & NI & I 
& \multicolumn{1}{c|}{NI} & NI & \multicolumn{1}{c|}{I} & NI & I 
& \multicolumn{1}{c|}{NI} & NI & \multicolumn{1}{c|}{I} & NI & I \\ \hline

\multicolumn{1}{|c|}{Block} & 
\multicolumn{1}{c|}{0.135} & 0.139 & \multicolumn{1}{c|}{0.126} & 0.138 & 0.118 
& \multicolumn{1}{c|}{0.103} & 0.106 & \multicolumn{1}{c|}{0.098} & 0.103 & 0.093 
& \multicolumn{1}{c|}{0.083} & 0.081 & \multicolumn{1}{c|}{0.077} & 0.085 & 0.075 \\

\multicolumn{1}{|c|}{Smooth} & 
\multicolumn{1}{c|}{0.075} & 0.081 & \multicolumn{1}{c|}{0.073} & 0.080 & 0.070 
& \multicolumn{1}{c|}{0.057} & 0.058 & \multicolumn{1}{c|}{0.058} & 0.055 & 0.057 
& \multicolumn{1}{c|}{0.041} & 0.041 & \multicolumn{1}{c|}{0.042} & 0.042 & 0.043 \\

\multicolumn{1}{|c|}{Hotspots} & 
\multicolumn{1}{c|}{0.135} & 0.137 & \multicolumn{1}{c|}{0.140} & 0.131 & 0.131 
& \multicolumn{1}{c|}{0.117} & 0.119 & \multicolumn{1}{c|}{0.119} & 0.117 & 0.118 
& \multicolumn{1}{c|}{0.109} & 0.110 & \multicolumn{1}{c|}{0.110} & 0.111 & 0.111 \\

\multicolumn{1}{|c|}{Random} & 
\multicolumn{1}{c|}{0.171} & 0.174 & \multicolumn{1}{c|}{0.180} & 0.175 & 0.175 
& \multicolumn{1}{c|}{0.152} & 0.144 & \multicolumn{1}{c|}{0.155} & 0.145 & 0.138 
& \multicolumn{1}{c|}{0.127} & 0.105 & \multicolumn{1}{c|}{0.116} & 0.109 & 0.102 \\ \hline

\end{tabular}
}
\end{table}

\begin{figure}[!t]
\centering
\includegraphics[scale=.6]{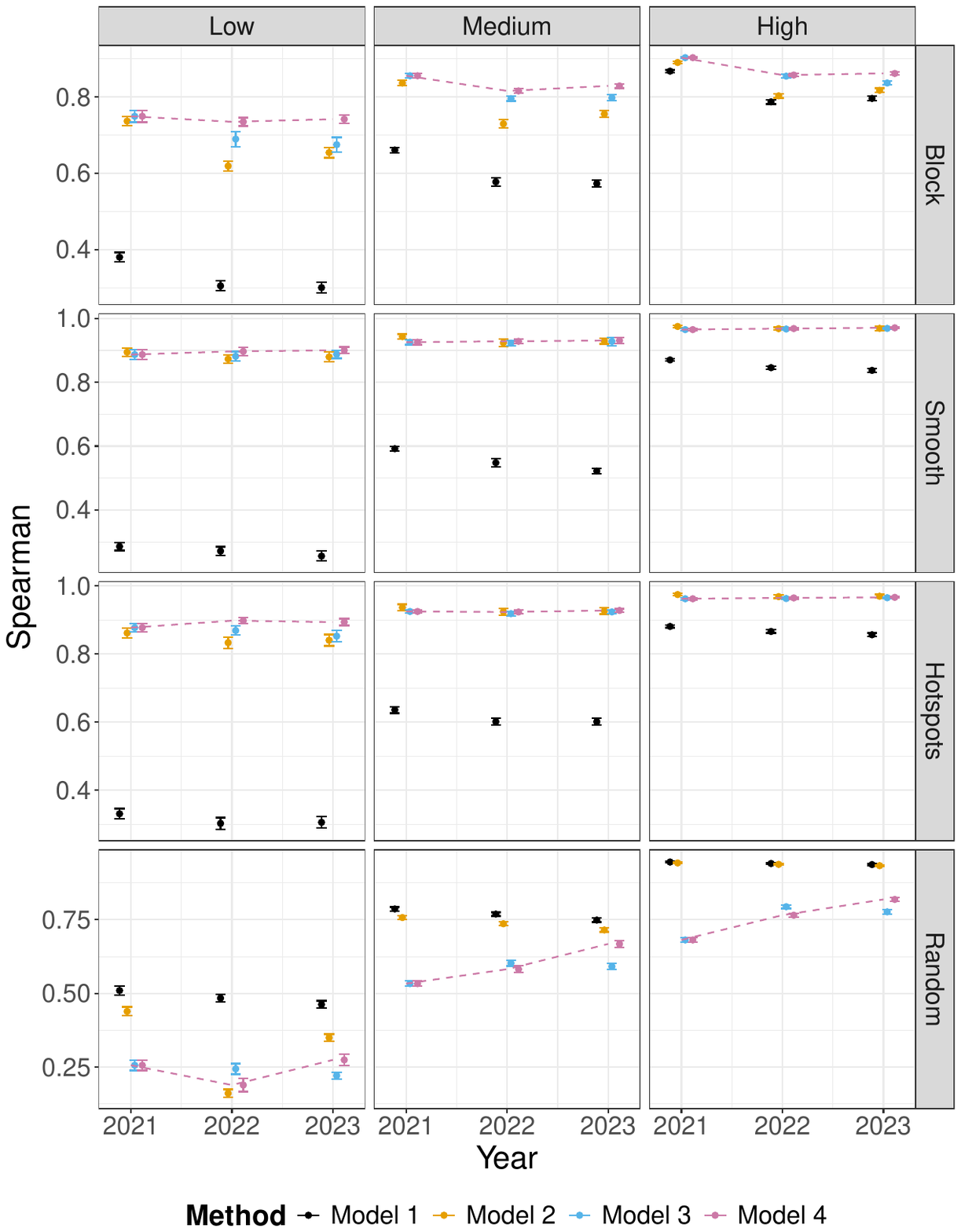}
\caption{\label{fig:spearman}Mean spearman correlation between true and estimated ZIP Code effects across spatial patterns, precision settings, and years. Error bars denote 95\% confidence intervals across the 30 replicates. Model 1 represents the frequentist beta regression model with non-spatial random effects. Model 2 represents the Bayesian beta regression with spatial random effects using an ICAR prior. Model 3 represents our proposed model without temporal borrowing and Model 4 represents the proposed model with temporal borrowing.}
\end{figure}

\section{Diagnostics for the Case study}

\begin{figure}[H]
\centering
\caption{Histogram of Gelman-Rubin $(\hat{R})$ statistic for all model parameters for each of the three years' models. Note that in 2022 and 2023, all $\hat{R}$ values are under 1.01 and in the 2021 model, all $\hat{R}$ values are under 1.012, which suggests good model convergence in all models.}
\resizebox{\textwidth}{!}{%
\begin{tabular}{|c|c|c|}
\hline
\subfigure[2021]{\label{fig:Rhat2021}\includegraphics[trim=0cm 0cm 0cm 0cm, clip, scale=0.33, page=1]{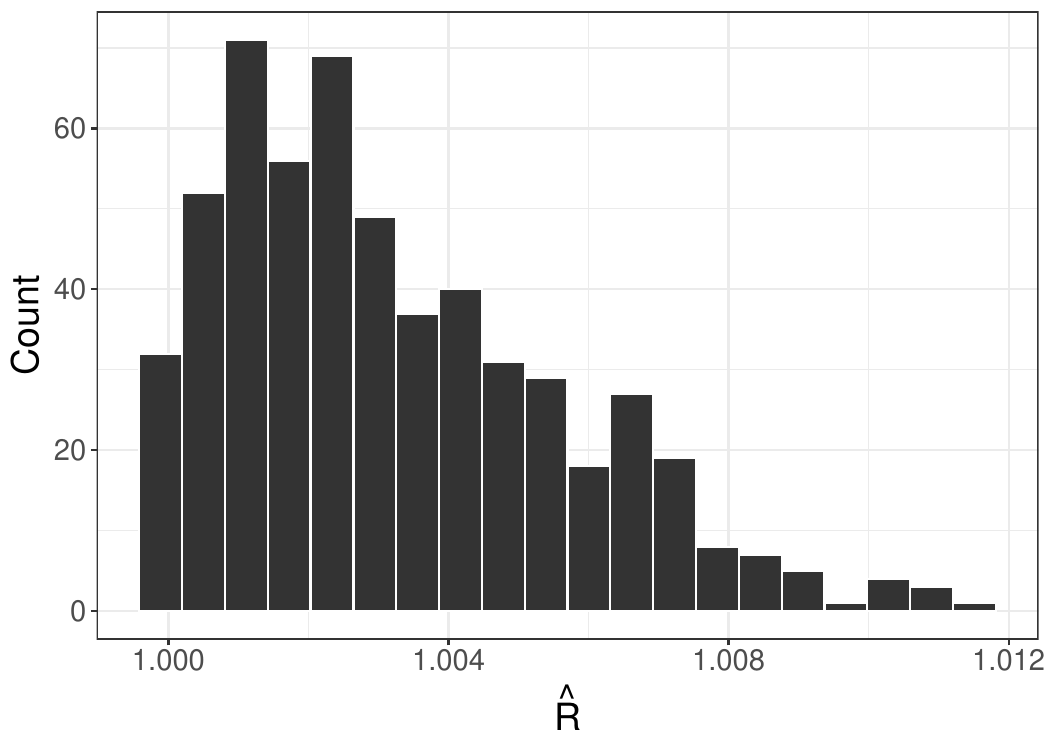}} & 
\subfigure[2022]{\label{fig:Rhat2022}\includegraphics[trim=0cm 0cm 0cm 0cm, clip, scale=0.33, page=1]{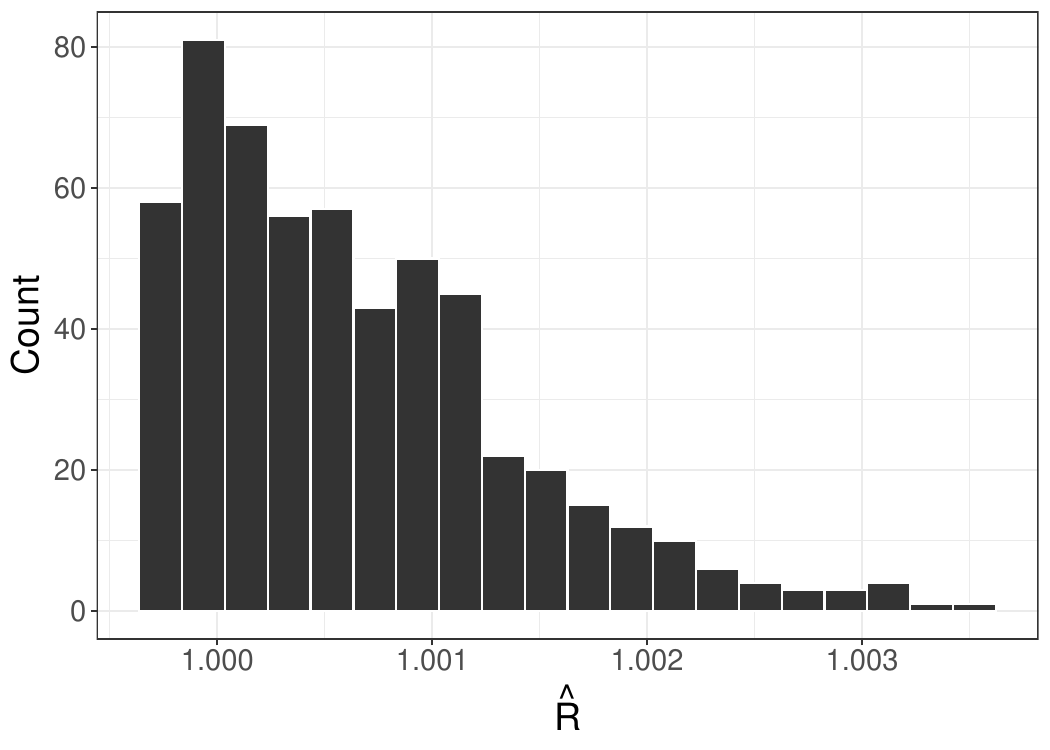}} &
\subfigure[2023]{\label{fig:Rhat2023}\includegraphics[trim=0cm 0cm 0cm 0cm, clip, scale=0.33, page=1]{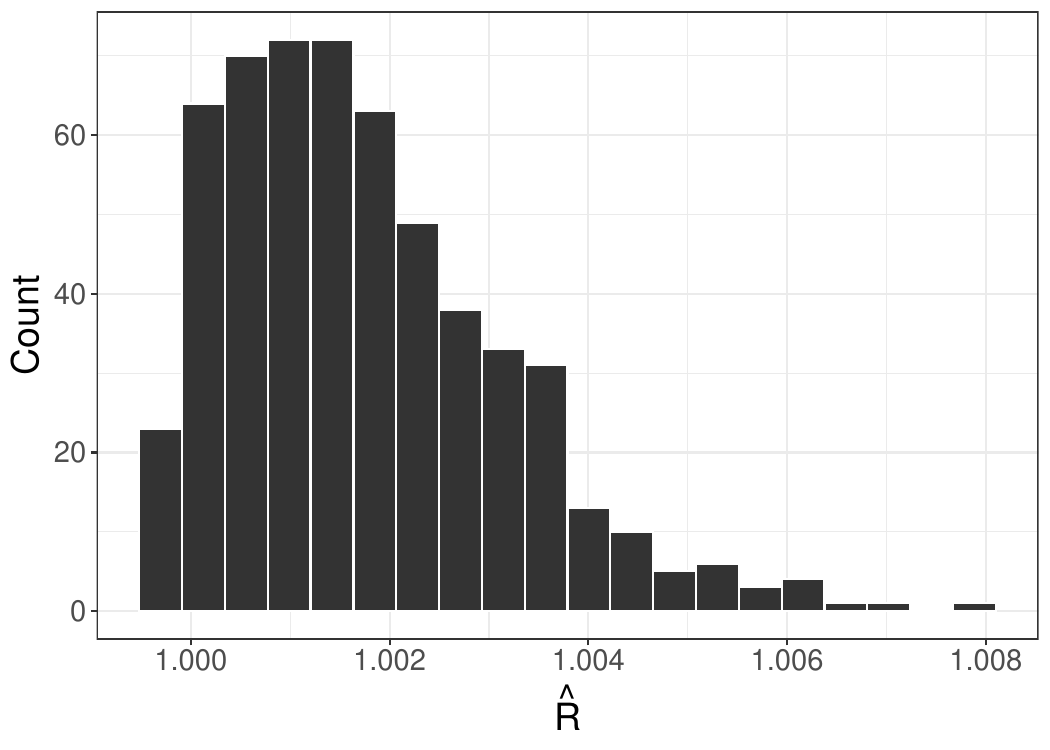}} \\ \hline
\end{tabular}
}
\end{figure}

\section{Computational performance}\label{subsec:time}
To assess the computational demands of our Bayesian beta regression model, we recorded the runtime for the MCMC sampling across all three years of analysis. Model estimation was performed using Stan, with each chain executed on a separate node of our institution's computing cluster via batch job submission. Each chain consisted of 4,000 warm-up iterations followed by 6,000 post-warm-up samples. In each year, the model specification included 14 covariates in the mean model and 6 covariates in the precision model. In 2021, the dataset comprised 4,443 individual-level observations. The average warm-up and sampling times per chain were 12.8 and 19.8 hours, respectively, yielding a total average runtime of 32.6 hours. For 2022, with 3,602 observations, the average warm-up and sampling times were 14.6 and 31.1 hours, respectively, resulting in a total runtime of 45.7 hours. In 2023, there were 3,428 observations and the average warm-up and sampling times were 10.8 hours and 16.5 hours, respectively, for a total average runtime of 27.3 hours. These results provide an estimate for the computational time of fully Bayesian estimation for our high-dimensional hierarchical modeling framework, particularly when incorporating spatially- and temporally-informed priors and modeling both mean and precision components.

\section{Stan and R codes that fit the double generalized Bayesian beta regression model}

\begin{lstlisting}[caption={Stan code for double generalized Bayesian beta regression model}, label={lst:stan}]
data {
  int<lower=1> n;                      // sample size
  int<lower=1> p;                      // p individual predictors (mean model)
  int<lower=1> q;                      // q zcta predictors (dispersion model)
  int<lower=1> S;                      // S zctas (all ZCTAs - with and without obs)
  vector<lower=0,upper=1>[n] y;        // response 
  matrix[n,p] X;                       // individual predictor matrix (mean model)
  matrix[S,q] C;                       // zcta predictor matrix (dispersion model)
  matrix[n,S] Z;                       // 0/1 zcta matrix (which ZCTA each obs belongs to)
  matrix[S,S] L;                       // graph Laplacian matrix
  vector[S] alphaPriorMean;            // vector of alpha priors (in t=1 it is 0 vector, in t>1, it is alpha.hat_t-1)
  matrix[S,S] I;                       // identity matrix
  real betaPriorMean;                  // prior means for beta coefficients (usually 0) (mean model)
  real<lower=0> betaPriorVar;          // prior variance for beta coefficients (noninformative) (mean model)
  real omegaPriorMean;                 // prior means for zcta level covariates (usually 0) (dispersion model)
  real<lower=0> omegaPriorVar;         // prior variance for zcta level covariates (noninformative) (dispersion model)
  real<lower=0> lambdaPrior1;          // shape parameter for lambda prior (lambda: tuning parameter for the Laplacian)
  real<lower=0> lambdaPrior2;          // rate parameter for lambda prior
  real<lower=0> gammaPrior1;           // shape parameter for gamma prior (gamma: tuning parameter for the ridge)
  real<lower=0> gammaPrior2;           // rate parameter for gamma prior 
}

parameters {
  vector[p] beta;                      // reg coefficients (mean model)
  vector[S] alpha;                     // zip-code effect (mean model)
  vector[q] omega;                     // reg coefficients (dispersion model)
  real<lower=0> lambda;                // lambda penalty
  real<lower=0> gamma;                 // gamma penalty
}

transformed parameters {
  matrix[S,S] M;
  M = lambda*L + lambda*gamma*I;       // inverse of M is covariance matrix
  matrix[S,S] Minv;
  Minv = inverse(M);                   // inverse of M
  vector[n] eta;                       // mean model
  vector[S] zeta;                      // dispersion model
  eta = X*beta + Z*alpha;              // individual level variables
  zeta = C*omega;                      // zcta level variables 
  
  vector[n] mu;                        // transformed linear predictor
  vector[S] phi;                       // dispersion for each ZCTA
  vector[n] phi2;                      // dispersion for each obs
  vector[n] A;                         // parameter for beta distn
  vector[n] B;                         // parameter for beta distn

  mu = inv_logit(eta);                 // mean model logit link

  phi = exp(zeta);                     // dispersion model log link

  phi2 = Z * phi;                      // reorganized dispersion (for all obs)

  A = mu .* phi2;                      // usual shape parameters (A and B)
  B = (1.0 - mu) .* phi2;
}

model {
  // priors
  beta ~ normal(betaPriorMean, betaPriorVar);
  omega ~ normal(omegaPriorMean, omegaPriorVar);
  alpha ~ multi_normal(alphaPriorMean, Minv); // alpha prior
  lambda ~ gamma(lambdaPrior1, lambdaPrior2); // lambda prior
  gamma ~ gamma(gammaPrior1, gammaPrior2); // gamma prior

  // likelihood
  y ~ beta(A, B);
}

\end{lstlisting}

\begin{lstlisting}[caption={R code to implement the Stan model}, label={lst:R}]
stan(file=".../betareg_spatial.stan",
     data = list(n = nrow(Xmat), p = ncol(Xmat), 
                 q = ncol(Cmat), S = nrow(Cmat), 
                 y = y, X = Xmat, C=Cmat, 
                 Z = Z, L = L, I = diag(nrow(Cmat)),
                 betaPriorMean = 0, betaPriorVar = 10,
                 omegaPriorMean = 0, omegaPriorVar = 10,
                 alphaPriorMean = Zero, lambdaPrior1 = 1,
                 lambdaPrior2 = 1, gammaPrior1 = 1, gammaPrior2=5),
     pars = c("beta","omega","alpha","lambda","gamma"),
     chains = 1,
     iter = n.iter,
     warmup = n.burnin,
     thin = n.thin,
     seed = seed2)
\end{lstlisting}

\end{document}